\documentclass[10pt, journal]{IEEEtran} 

\usepackage{cite}

\usepackage{siunitx}

\ifCLASSINFOpdf
  \usepackage[pdftex]{graphicx}
\else
  \usepackage[dvips]{graphicx}
\fi

\usepackage[cmex10]{amsmath}
\usepackage{amssymb}
\usepackage{upgreek}

\usepackage[tight,footnotesize]{subfigure}

\usepackage{url}

\usepackage[pdftex,dvipsnames]{xcolor}
\usepackage{todonotes}
\definecolor{morange}{rgb}{0.8,0.2,0}
\definecolor{mblue}{rgb}{0,0.3,1.0}
\definecolor{mpink}{rgb}{1.0,0.6,0.6}

\newcommand{\invitro}{\emph{in vitro}}

\newcommand{\stepsize}[1]{\Delta r_{#1}}
\newcommand{\xth}[1]{ {#1}\textrm{th} }

\newcommand{\vavg}{v_{\text{avg}}}
\newcommand{\perslen}{L_p}

\begin{document}
\title{A Comprehensive Survey of Recent \\ Advancements in Molecular Communication}

\author{Nariman~Farsad,~\IEEEmembership{Member,~IEEE,}
			H. Birkan~Yilmaz,~\IEEEmembership{Member,~IEEE,}
			Andrew~Eckford,~\IEEEmembership{Senior Member,~IEEE,}
        Chan-Byoung~Chae,~\IEEEmembership{Senior Member,~IEEE,}
        and~Weisi Guo,~\IEEEmembership{Member,~IEEE}
\thanks{Manuscript received September 2, 2015; revised December 21, 2015; accepted February 5, 2016}%
\thanks{N.~Farsad is with the Department of Electrical Engineering, Stanford University, CA, USA. e-mail: nfarsad@stanford.edu}%
\thanks{H. B.~Yilmaz and C.-B.~Chae are with the School of Integrated Technology, Institute of Convergence Technology, Yonsei University, Korea. e-mails: \{birkan.yilmaz, cbchae\}@yonsei.ac.kr}%
\thanks{A.~W.~Eckford is with the Department of Electrical Engineering and Computer Science, York University, Ontario, Canada. e-mail: aeckford@cse.yorku.ca}%
\thanks{W. Guo is with the School of Engineering, University of Warwick, Coventry, UK. e-mail: weisi.guo@warwick.ac.uk}
\thanks{The work of H. B. Yilmaz and C.-B. Chae was in part supported by the
MSIP, under the ``IT Consilience Creative Program" (IITP-2015-R0346-15-1008) and by the Basic Science Research Program (2014R1A1A1002186), through the NRF of Korea.}}
\maketitle


\normalsize

\begin{abstract}
With much advancement in the field of nanotechnology, bioengineering and synthetic biology over the past decade, microscales and nanoscales devices are becoming a reality. Yet the problem of engineering a reliable communication system between tiny devices is still an open problem. At the same time, despite the prevalence of radio communication, there are still areas where traditional electromagnetic waves find it difficult or expensive to reach. Points of interest in industry, cities, and medical applications often lie in embedded and entrenched areas, accessible only by ventricles at scales too small for conventional radio waves and microwaves, or they are located in such a way that directional high frequency systems are ineffective. Inspired by nature, one solution to these problems is molecular communication (MC), where chemical signals are used to transfer information. Although biologists have studied MC for decades, it has only been researched for roughly 10 year from a communication engineering lens. Significant number of papers have been published to date, but owing to the need for interdisciplinary work, much of the results are preliminary. In this paper, the recent advancements in the field of MC engineering are highlighted. First, the biological, chemical, and physical processes used by an MC system are discussed. This includes different components of the MC transmitter and receiver, as well as the propagation and transport mechanisms. Then, a comprehensive survey of some of the recent works on MC through a communication engineering lens is provided. The paper ends with a technology readiness analysis of MC and future research directions.
\end{abstract}


\section{Introduction}
\label{sec:intro}
\IEEEPARstart{T}{he} problem of conveying information over a distance has always been an important part of human society. Today, modern communication systems solve this problem with electrical or electromagnetic signals from radio to optical bands. There are, however, still many applications where these technologies are not convenient or appropriate. For example, the use of electromagnetic wireless communication inside networks of tunnels, pipelines, or  salt water environments, can be very inefficient\cite{eckBook,guo2015molecularVE,stojanovic2003acousticUC}. As an example, electromagnetic waves do not propagate effectively over long distances through the ocean. The salinity of sea water causes conductivity, which leads to rapid attenuation of electromagnetic signals, especially at higher frequencies\cite{stojanovic2003acousticUC}. As another example, at extremely small dimensions, such as between microscale or nanoscale robots \cite{freitas-book, sen12}, electromagnetic communication is challenging because of constraints such as the ratio of the antenna size to the wavelength of the electromagnetic signal~\cite{aky08}. Optical communication is also not suitable for these applications since it requires either a guided medium (e.g. fiber optical cable) or line of sight. 

Inspired by nature, one possible solution to these problems is to use {\em chemical signals} as carriers of information, which is called {\em molecular communication} (MC) \cite{eckBook}. The reliability of the molecular signal in terms of reaching the intended receiver can be greater than radio-based communications in challenging environments where the radio signals experience heavy diffraction loss~\cite{guo2015molecularVE}. In MC, a transmitter releases small particles such as molecules or lipid vesicles into an aqueous or gaseous medium, where the particles propagate until they arrive at a receiver. The receiver then detects and decodes the information encoded in these particles. 

In nature, chemical signals are used for inter-cellular and intra-cellular communication at microscales and nanoscales \cite{alb07}, while pheromones are used for long range communication between members of the same species such as social insects \cite{ago92-book}. Therefore, chemical signals can be used for communication at both macroscopic and microscopic scales. Moreover, MC signals are biocompatible, and require very little energy to generate and propagate. These properties makes chemical signals ideal for many applications, where the use of electromagnetic signals are not possible or not desirable. Although MC is present in nature and is used by microorganisms such as microbes to communicate and detect other microorganisms, it was only recently that engineering an MC system has been proposed as means of communication at the microscale \cite{hiy05}. Macroscale MC was not even considered until \cite{far13}.
\begin{figure*}[t]
	\begin{center}
		\includegraphics[width=2.00\columnwidth,keepaspectratio] %
		{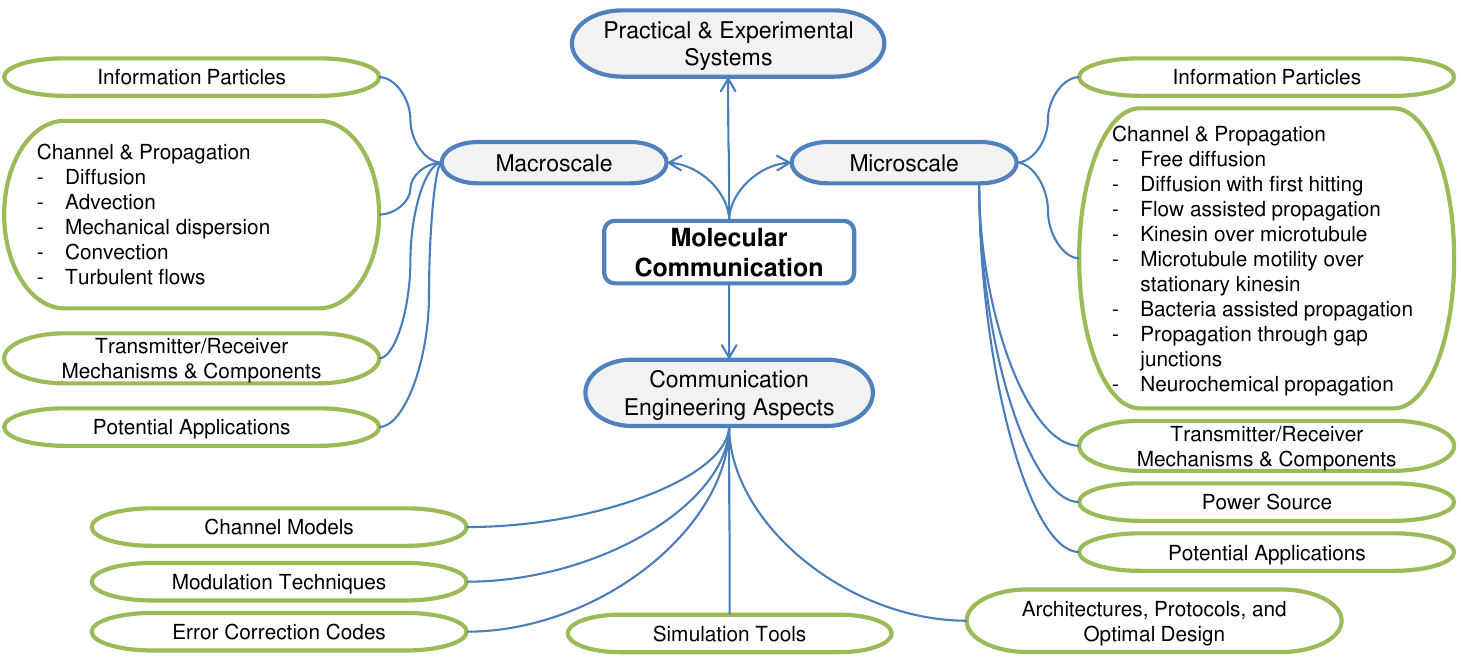}
	\end{center}
	\caption{\label{fig:variousAspects} {Various aspects of molecular communication.}}
\end{figure*}

In this paper we present a comprehensive survey of recent advancements of MC in a tutorial fashion. Some of the previous survey papers on MC are as follows. One of the early surveys that considered the concept of nanonetworks was \cite{aky08}, where both MC and electromagnetic-based communication using novel materials such as carbon nanotubes are considered. Long-range MC propagation schemes were proposed in \cite{gin09}. The first survey work on general microscale MC was presented by Hiyama and Moritani in \cite{hiy10NanoCom}. A more recent general microscale survey was presented by Nakano \textit{et al.} in \cite{nak12NBS}. In \cite{dar13}, a survey of MC based on microtubules and physical contact was presented, and \cite{bal13NanoCom} provided a guide post for some of the experimental problems within MC. Finally, a survey of MC from a layered communication network perspective was recently presented in \cite{nak14TNB}. There are also some magazine articles on specific areas of MC: in \cite{aky11} the concept of nanonetworks was presented; in~\cite{guo2015molecularCC} the physical layer techniques and channel models were presented; an overview of communication through gap junction channels  was presented in \cite{kur12WCM}; \cite{yeh14} discussed diffusion-based MC; and the concept of intra-body nervous nanonetworks was presented in \cite{mal14}. There are also a number of books on MC at microscales \cite{bush-book,eckBook,ataBook}. 

This paper is different from all these works in the following ways. First, we provide a more general description of MC, one which considers MC at both macroscales and microscales, simultaneously. Second, we provide a comprehensive survey of MC at both scales. All of the previous survey work have between 60-100 references, since they provide a survey of a typical sub-field of MC. In this work, we provide over 200 citations (some of these references are also cited in other surveys) many of which are very recent works (including some works from early 2015). Third, many new advancements have been highlighted here which have been achieved since the publication of these previous surveys. Just like many of these great surveys, our paper is presented in a tutorial fashion such that the researchers new to the field could use it to familiarize themselves quickly with some of the fundamental underlying physical and chemical processes and mathematical models. Finally, most content provided in this survey focuses on the physical and the link layer design.

In this paper, we elaborate on various aspects of MC with focusing recent advancements and Figure~\ref{fig:variousAspects} summarizes the content of this paper. In Section~\ref{sec:overview}, an overview of MC is provided. Section~\ref{sec:MicroMC} describes the underlying biological components, as well as chemical and physical processes that are employed in engineering MC systems at microscales, while Section~\ref{sec:MacroMC} considers macroscale MC systems. In Section~\ref{sec:ComEngMC}, recent advancements in communication engineering aspect of MC is presented. Section~\ref{sec:pracSys} highlights some of the practical and experimental setups that demonstrate the feasibility of engineering MC systems. Finally, concluding remarks are presented in Section~\ref{sec:conc}.

\section{Overview of Molecular Communication}
\label{sec:overview}

\begin{figure}[t]
	\begin{center}
		\includegraphics[width=0.98\columnwidth,keepaspectratio] %
		{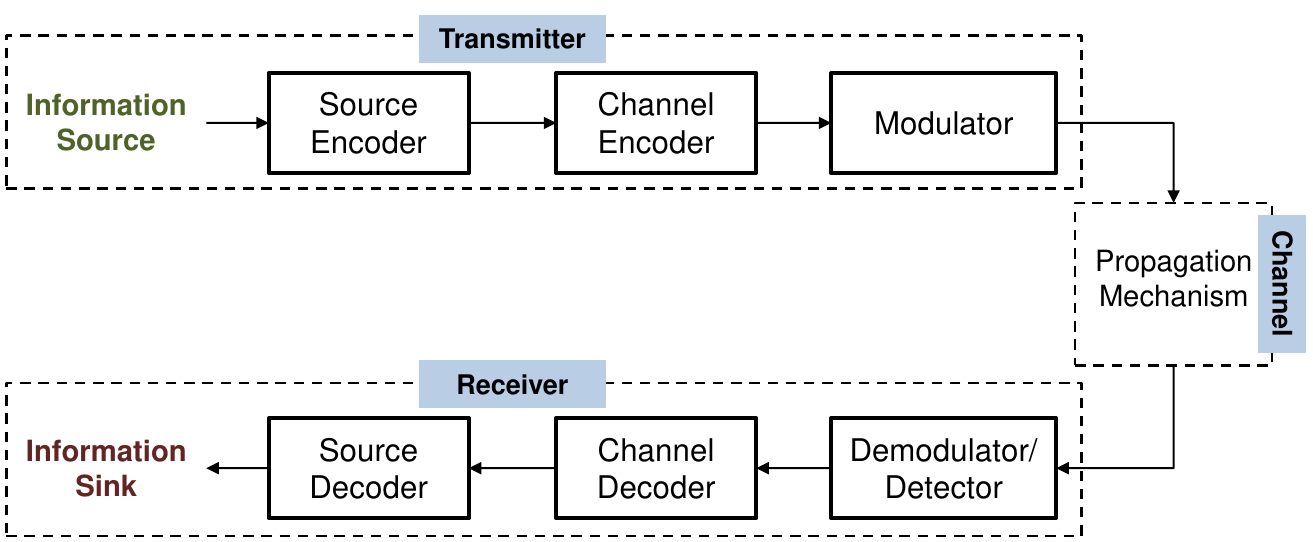}
	\end{center}
	\caption{\label{fig:commDiag} {Block diagram of a typical communication system.}}
\end{figure}

The goal of a communication system is to transfer information across space and time. To achieve this, a signal needs to be generated by a transmitter in accordance to the information that is intended to be transferred to a receiver. This signal then propagates to a receiver, where the intended information is decoded from the received signal. Therefore,  any communication system can be broken down into three major components: the transmitter, the receiver, and the channel. Figure \ref{fig:commDiag} shows the block diagram representation of these three modules and their submodules. 

%

The channel is the environment in which the transmitted signal propagates from the sender to the receiver. In a traditional communication system, a channel is typically a wire or free space where the transmitted signals are the electrical currents or electromagnetic waves, respectively. In MC, small particles called {\em information particles} act as chemical signals conveying the information. Information particles are typically a few nanometers to a few micrometers in size. Information particles could be biological compounds, such as proteins, or synthetic compounds, such as gold nano particles. The channel in MC is an aqueous or a gaseous environment where the tiny information particles can freely propagate.     

In communication systems, the channel may introduce noise into the system, where the noise is any distortion that results in degradation of the signal at the receiver. 
In radio-based communication systems the source of the noise is typically the fading of electromagnetic signals, and the interference of different electromagnetic waves, for example due to multipath. In MC the sources of the noise can be the followings
\begin{itemize}
\item random propagation (diffusion) noise
\item transmitter emission noise
\item receiver counting/reception noise
\item environment noise such as degradation and/or reaction
\item multiple transmitters
\item etc.
\end{itemize}

When the transmitted signal arrives at the receiver, the receiver must first demodulate and detect the channel symbols. The estimated channel symbols are then decoded using a channel decoder, where some of the errors introduced by the transmitter, the channel and the receiver may be corrected. The output of the channel decoder goes through a source decoder, where the receiver estimates what information the transmitter has sent. If this estimation is correct, then the communication session has been successful.

\begin{figure}[t]
	\begin{center}
		\includegraphics[width=3.4in] %
		{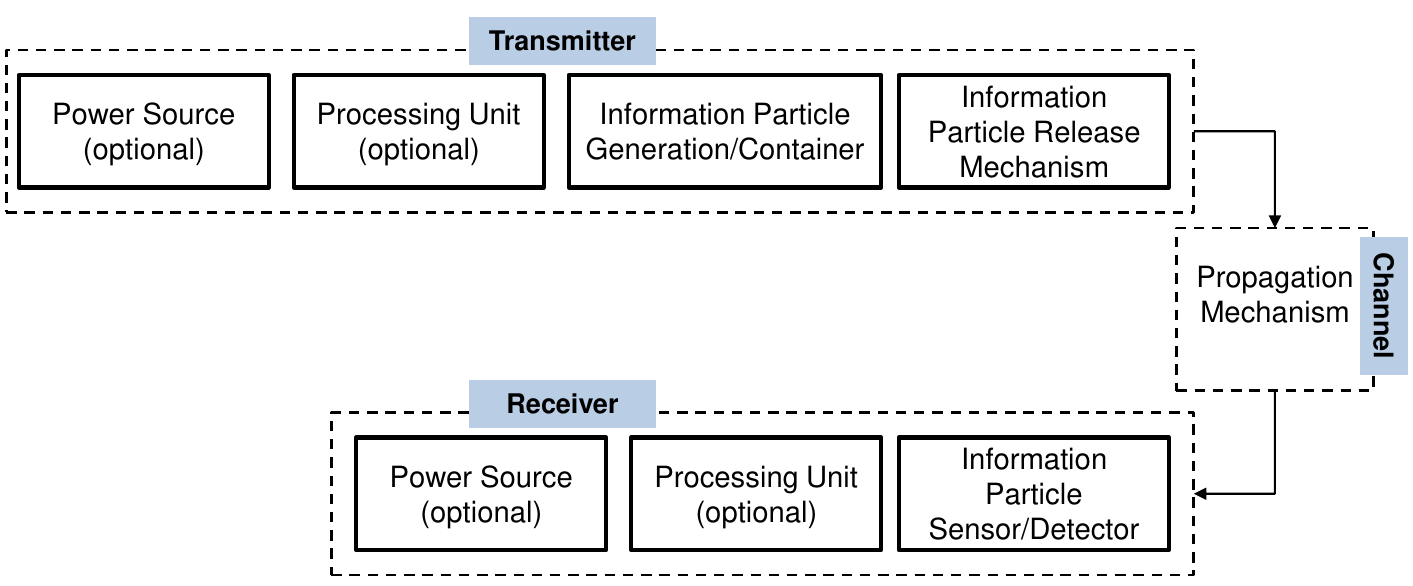}
	\end{center}
	\caption{\label{fig:MCcomp} {Physical components required for MC.}}
\end{figure}

Some of the physical components that may be required to implement MC systems are shown in Figure \ref{fig:MCcomp}. At the transmitter, a physical process is required for generation or storage of information particles.
There may also be a need for a mechanism that controls the release of information particles. Finally, there should be a processing unit that controls the different processes within the transmitter. The processing unit can operate chemically (e.g., chemical pathways), electrically (e.g., microcontroller), or through other means. To function, the transmitter may need a source of power. For example, the transmitter can be a synthetic cell that gets its power from the environment or from an electrical source. 

After the information-carrying particles are released in the channel, a propagation mechanism is necessary for transporting them to the receiver. This mechanism can be diffusion based, flow based, or an engineered transport system using molecular motors. At the receiver, there must be a sensor, receptor, or a detector that can measure some arrival property of the received information particles. This property can be simply the presence or absence of the information particles, their concentration, time of arrival, or any other measurable parameter. If coding techniques are used there may also be a need for a central processing unit for decoding and deciphering the received signal. Finally, a power source may be required for the receiver to function.

In nature, MC is employed over short-range (nm scale), mid-range ($\upmu$m to cm scale), or long-range (cm to m scale) communication \cite{alb09-book, pol07-book}. For example, neurotransmitters use passive transport (free diffusion) to communicate over short-range; inside cells, motor proteins are used to actively transport cargoes over the mid-range; and hormones are transported over the long-range using flow (e.g., blood flow from the heart). In this survey we refer to the short- and mid-range as microscale MC, and the long-range as macroscale MC. The physical properties of matter change from macroscales to microscales, hence we consider each of them separately in the next two subsection, and discuss different mechanisms that can be implemented at each scale.

\section{Microscale Molecular Communication}
\label{sec:MicroMC}

On December 29, 1959, physicist Richard Feynman gave a lecture at American Physical Society meeting in Caltech titled ``There's Plenty of Room at the Bottom'' \cite{fey59}. The goal of the talk was to inspire scientists for finding a solution to ``the problem of manipulating and controlling things on a small scale''. To motivate his vision of a never ending limits of miniaturization he asked ``Why cannot we write the entire 24 volumes of the Encyclopaedia Brittanica on the head of a pin?'' 

Over the past couple of decades there have been considerable advancements in the fields of nanotechnology, biotechnology, and microrobotics, where designing and engineering of {\em microscale or nanoscale devices}
begin to take shape. In order to achieve complex goals, cooperation of these devices is essential, therefore a microscale or nanoscale networks \cite{aky08,bush-book} (typically called {\em nanonetworks}) of multiple micro- or nano-devices must be formed. This leads to a need for miniaturization of current communication systems, and  poses a question similar to the one Feynman asked half a century ago: how much can we shrink a communication system?  



At small scales, two prominent modes of communication are considered: miniaturizing the traditional electronic or electromagnetic communication systems and utilizing/mimicking the existing MC systems in nature. One advantage of miniaturizing the traditional electronic or electromagnetic communication is the availability of a rich theory that communication engineers have developed over the past century. For molecular commutation, a comprehensive theoretical framework is not known to communication engineers, however, fully functional MC systems have already been selected by evolutionary processes, and used by living organisms over billions of years. 

MC has two main advantages over the electromagnetic communication for nanonetworks. First, it is biocompatible, which means that it could be used for medical applications with ease. Second, it is  energy efficient and has a very low heat dispersion. Therefore, MC becomes a more suitable solution for nanonetworks.

Over the next subsections we describe the different physical, chemical and biological processes that underlie  MC systems. Since one of the main differences  between MC and radio based communication is the carrier signal, we start by presenting information particles and their propagation.

\subsection{Information Particles}

The structure and size of these information particles may affect how they propagate in the environment. For example, increasing the size of the information particle changes the diffusion coefficient and hence the diffusion process.  To make the MC channel more reliable, information particles need to be chemically stable and robust against environmental noise and interference from other molecules and particles. Moreover, information particles may degrade or denature over time in the  environment. For example, molecular deformation and cleavage caused by enzymatic attacks or changes in pH in the environment may degrade the information particles \cite{hiy10NanoCom}.

\begin{figure}[ht]
	\begin{center}
		\includegraphics[width=3.4in] %
		{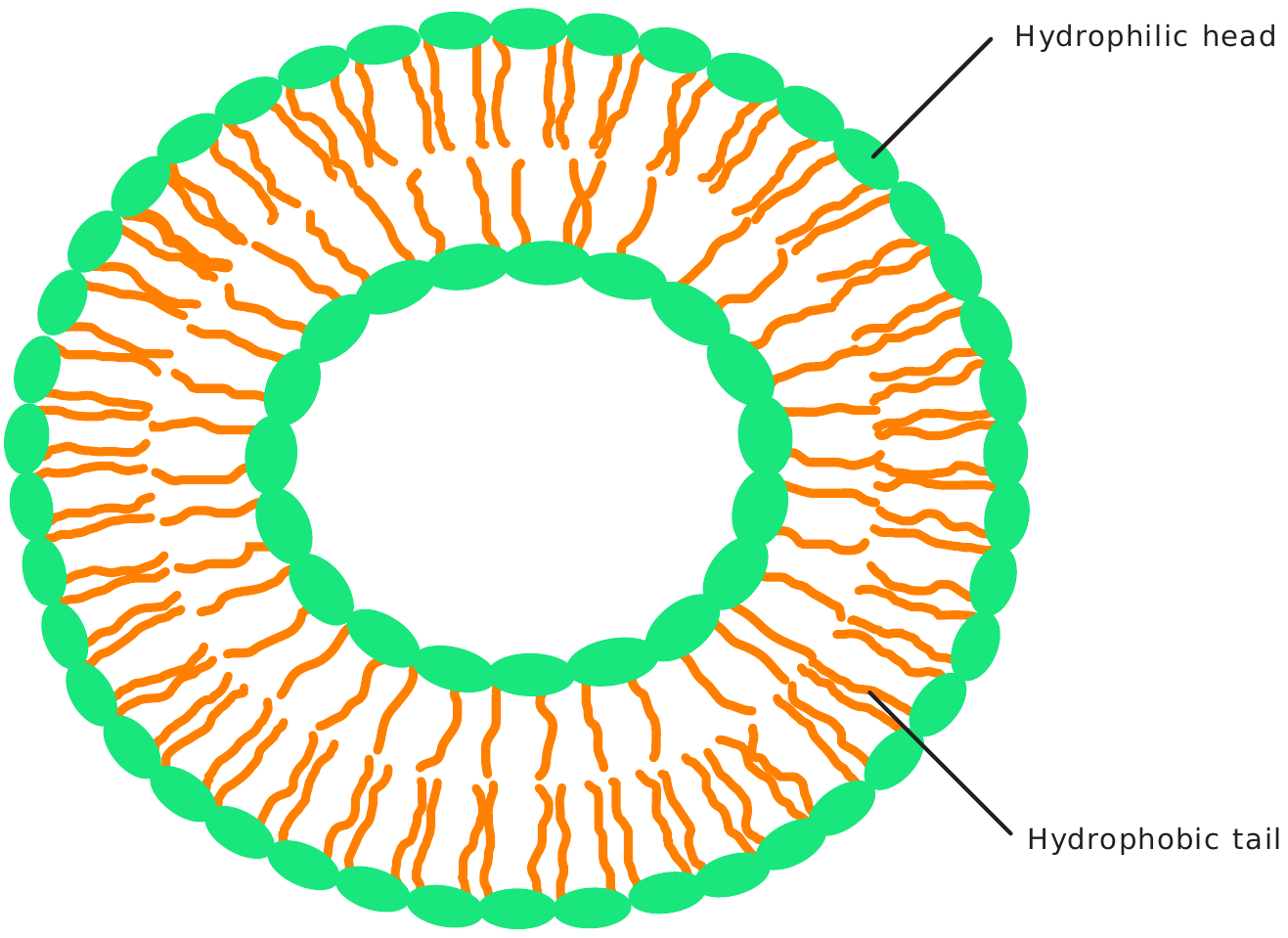}
	\end{center}
	\caption{\label{fig:liposome} {Liposome's lipid bilayer structure.}}
\end{figure}

Due to degradation of particles, the transmitter can release a large number of information particles, or alternatively encapsulate the information-carrying particles inside {\em liposomes} that separate the environment and the information particles.  Liposomes are lipid bilayer structures that can encapsulate different proteins and particles \cite{nag00}. Figure \ref{fig:liposome} shows the structure of the liposome. It contains two layers of lipid molecules, where one side of the lipid is hydrophobic (repelled from water) and the other side is hydrophilic (attracted towards water). It separates the aqueous environment inside the liposome from the outside of the liposome. In \cite{hua1978}, the relationship between the number of lipid molecules and the liposome's wall thickness and radius is derived and tested experimentally. In \cite{mat11}, a chemical assembly line for generation of lipid vesicles and encapsulation process is presented.

Some examples of information particles used in nature by biological systems include hormones, pheromones, neurotransmitters, intracellular messengers, and deoxyribonucleic acid (DNA) and ribonucleic acid (RNA) molecules. Information particles can also be synthesized for specific purposes as demonstrated in drug delivery. For example, in \cite{lav03} nanoparticles are used to target particular tissue types.  



\subsection{Channel and Propagation}
\label{subsec:MicroProp}
For microscale MC, many different propagation schemes are possible for transporting information particles. They are:
diffusion based propagation \cite{nak08,pie10, pie13}, flow assisted propagation \cite{sri12}, active transport using molecular motors and cytoskeletal filaments \cite{hiy10LabChip}, bacterial assisted propagation \cite{gre10, cob10, lio12}, and kinesin molecular motors moving over immobilized microtubule (MT) tracks \cite{moo09b,eno11}. We describe each propagation method in detail over the next few subsections and  compare the propagation schemes in Table~\ref{tb:micro_propagation_comparison} at the end of this subsection.

\subsubsection{Free Diffusion}
Diffusion, also known as Brownian motion, refers to the random motion of a particle as it collides with other molecules in its vicinity. Through this random motion, the information-carrying particles can propagate from the transmitter to the receiver by utilizing the thermal energy that is already present in the channel environment. Therefore, no source of external energy is required for diffusion-based propagation. There are many examples of this class of propagation in biology. An example is the neurotransmitter molecules Acetylcholine (ACh) emitted from nerve cell to neuromuscular junction for conveying motor actions in muscle cells \cite{kuran2013aTunnelBA}. When muscles in a specific part of the body need to contract, the nerve cells in that region send a signal to the muscle tissue through these junctions to trigger  contraction \cite{alb07}. Another example is DNA binding molecules (e.g., repressors) propagating over a DNA segment to search for a binding site~\cite{ber93-book}. 

Diffusion can be accurately simulated using Monte Carlo simulation \cite{ber-book}. In particular, the motion of the information particles is considered for a discrete time period of duration $\Delta t$. Given some initial position $(x_0,y_0,z_0)$ at time $t = 0$, for any integer $k > 0$, the motion of the particles is given by the sequence of coordinates $(x_i,y_i,z_i)$ for $i=1, 2, \ldots, k$. Each coordinate $(x_i,y_i,z_i)$ represents the position of the particle at the time $t = i \Delta t$
\begin{eqnarray}
	\label{eqn:bBMXYZCoord}
	(x_i, y_i, z_i) & = & (x_{i-1}, y_{i-1}, z_{i-1}) + (\Delta x_i, \Delta y_i, \Delta z_i)\\
	\label{eqn:bBMXCoord}
	\Delta x_i & \sim & \mathcal{N}(0, 2D\Delta t) , \\
	\label{eqn:bBMYCoord}
	\Delta y_i & \sim & \mathcal{N}(0, 2D\Delta t) , \\
	\label{eqn:bBMZCoord}
	\Delta z_i & \sim & \mathcal{N}(0, 2D\Delta t) ,
\end{eqnarray}
where $\Delta x_i$, $\Delta y_i$, $\Delta z_i$, $D$ and $\mathcal{N}(\mu, \sigma^2)$ are the random displacements over each spatial axis during $\Delta t$, the diffusion coefficient, and the normal distribution with mean $\mu$ and the variance $\sigma^2$. The diffusion movement in these equations is a Martingale process, where subsequent motions are unrelated to previous motions and positions \cite{oksBook}.

The comparative sizes between the propagating molecule $S_m$ and the molecules of the fluid $S_{\text{fluid}}$ affect the diffusion coefficient \cite{tyrrell1984diffusionIL}. For a given particle and fluid environment, $D$ is given by
\begin{equation}
	\label{eqn:DiffConst}
	D = 
    \left\{ 
    	\begin{array}{ll}
    	\displaystyle\frac{k_B T}{6 \pi \eta R_H} & \text{if } S_m \gg S_{\text{fluid}}\\
        \vspace{0.01cm}    &  \\
        \displaystyle\frac{k_B T}{4 \pi \eta R_H} & \text{if } S_m \approx S_{\text{fluid}}
        \end{array}
    \right.,
\end{equation}
where $k_B = 1.38 \cdot 10^{-23}$ J/K is the Boltzman constant, $T$ is the temperature (in K), $\eta$ is the dynamic viscosity of the fluid, and $R_H$ is the hydraulic radius (also known as Stoke's radius) of the information particle. For most applications, it can be assumed that $D$ is stationary throughout the medium, and that collisions with the boundaries are elastic. Example values of $D$ in $\mu$m$^2$/s in water at 25$\,^\circ$C are obtained from \cite{bionumbersWeb} and listed in Table~\ref{tab:difcoeff}.
\begin{table}[ht]
\caption{Diffusion coefficients of selected molecules in water at 25$^\circ$ C.}
\label{tab:difcoeff}
\centering
\begin{tabular}{ll}
\hline  
Molecule   & $D$ ($\mu$m$^2$/s)                         \\
\hline 
DNA								& 0.81 to 53\\
Human serum albumin             & 61   		\\
Insuline 						& 150		\\
Sucrose							& 520 		\\
Glucose 						& 600  		\\
Glycerol 						& 930  		\\
Nitrate 						& 1700		\\
Water molecule					& 2100 		\\
\hline
\end{tabular}
\end{table}

\subsubsection{Diffusion with First Hitting}

In nature, most receptors remove the information molecules from the environment through binding, absorbing, or chemical reactions~\cite{cuatrecasas1974membraneR}. Therefore, in most cases almost all the molecules contribute to the signal at most once. Sometimes, instead of removing the molecules, biological systems have other mechanisms to guarantee that each molecule contributes to the signal only once (e.g., acetylcholinesterase breaks down the molecules in neuromuscular junctions) \cite{kuran2013aTunnelBA}. This process is modeled by first hitting processes since molecules are assumed to be removed from the environment by the receptors or the detection mechanisms.

For reception process, the nanonetworking literature has mostly considered an absorbing receiver in ${\text{1-dimensional}}$ (1-D) environment \cite{eck07}  or a hypothetical sphere without absorption in a 3-D medium~\cite{pie13}. First hitting formulations are handy when the receptors, detectors, or sensors detect the information particles by removing them from the environment or transforming them.

The first generalized model, in the nanonetworking domain, including first hitting process in 1-D environment was derived in~\cite{eck07}  as
\begin{align}
\label{eqn_fhit_pdf_1d}
f_\text{hit}^{1\text{D}} (t)= \frac{d}{\sqrt{4\pi D t^3\,} } \, e^{-d^2/4Dt}
\end{align}
where $d$ and $D$ correspond to distance and the diffusion coefficient, respectively. This equation can be interpreted as the impulse response of the diffusion channel in 1-D environment with an absorbing receiver. To find the probability of hitting an absorbing receiver until time $t$,  \eqref{eqn_fhit_pdf_1d} is integrated with respect to time 
\begin{align}
\begin{split}
F_\text{hit}^{1\text{D}} (t)\!=\! \int\limits_{0}^{t} f_\text{hit}(t^\prime)\,dt^\prime &=\! \text{erfc} \left( \frac{d}{\sqrt{4Dt\,}}\right)\\
 &=\! 2 \Phi\left(\frac{-d}{\sqrt{2Dt}}\right)
\end{split}
\end{align}
where $\text{erfc}(.)$ and $\Phi(.)$ represent the complementary error function and the standard Gaussian cdf, respectively.

Similarly, fraction of hitting molecules to a perfectly absorbing spherical receiver in a 3-D environment is derived in \cite{yilmaz20143dChannelCF,Akkaya_CL}. Hitting rate of molecules to a spherical receiver in a 3-D environment is formulated as 
\begin{align}
\label{eqn_fhit_pdf_3d}
f_\text{hit}^{3\text{D}}(t)= \frac{r_r}{d+r_r}\frac{d}{\sqrt{4\pi D t^3\,} } \, e^{-d^2/4Dt}
\end{align}
where $r_r$ denotes the radius of the spherical receiver~\cite{yilmaz20143dChannelCF}. One can obtain the fraction of hitting molecules until time $t$ by integrating $f_\text{hit} ^{3\text{D}}(t)$ in \eqref{eqn_fhit_pdf_3d} with respect to time, which yields similar results with the 1-D case. Note that, there is a positive probability of not hitting to the absorbing boundary  for a diffusing particle in a 3-D environment when time goes to infinity. The survival probability depends on the radius of the receiver and the distance. 

First hitting formulations in a 2-D environment with infinite boundaries do not exist. We just have the asymptotic behavior from \cite{redner2001guideTF}. In \cite{dy2008firstPT}, the analytical expressions are derived on a planar wedge with special wedge angle values. Authors derived closed form expressions for the wedge angles of $\pi/2$ and $\pi/k$ where $k$ is an odd integer.

\subsubsection{Flow Assisted Propagation}

Although diffusion is advantageous because no external energy is required for propagation, it can be very slow at transporting information particles from the transmitter to the receiver over large separation distances. One way to assist the speed of propagation in pure diffusion is to introduce flow into the environment. The most effective flow is the one that is in the direction of the the transmitter to the receiver. 
An example of this propagation scheme is found in biology. Certain cells can secrete hormonal substances that propagate using the flow present in the blood stream to distant target cells. 

The flow assisted propagation can be simulated using Monte Carlo simulations by modifying the equation presented in the previous section. Introducing flow will change Equations (\ref{eqn:bBMXCoord})-(\ref{eqn:bBMZCoord}) into,
%
%
%
\begin{align}
	\label{eqn:BMFXCoord}
	\Delta x_i  = &v_{x,i-1} (x_{i-1},y_{i-1},z_{i-1}) \Delta t +\mathcal{N}(0, 2D\Delta t), \\
	\label{eqn:BMFYCoord}
	\Delta y_i  = &v_{y,i-1} (x_{i-1},y_{i-1},z_{i-1}) \Delta t +\mathcal{N}(0, 2D\Delta t), \\
	\label{eqn:BMFZCoord}
	\Delta z_i  = &v_{z,i-1} (x_{i-1},y_{i-1},z_{i-1}) \Delta t +\mathcal{N}(0, 2D\Delta t) ,
\end{align}
where $v_{x,i-1} (x_{i-1},y_{i-1},z_{i-1})$, $v_{y,i-1} (x_{i-1},y_{i-1},z_{i-1})$, and $v_{z,i-1} (x_{i-1},y_{i-1},z_{i-1})$ are flow velocities in the $x$, $y$, and $z$ directions \cite{ber-book}. These velocities are function of time and space because in practice flows can change with respect to these variables.

\subsubsection{Motor Protein Moving Over Microtubule Tracks}

It is also possible to transport the information particles actively from the transmitter to the receiver using motor proteins. One way to achieve this is using microtubules and kinesin motor proteins \cite{alb97-book,hiy09,hiy10LabChip,eno11}.  Microtubules are hollow tube-like structures, 24 nm in diameter, whose walls are formed by adjacent protofilaments. They are made up of dimeric subunits composed of $\alpha$- and $\beta$-tubulin that polymerize into microtubules. In biology, microtubules are a component of the cytoskeleton, found throughout the cytoplasm. They are involved in maintaining the structure of the cell, providing platforms for intracellular transport, and number of other intracellular processes.

\begin{figure}[!t]
	\begin{center}
		\includegraphics[width=3.2in]
        {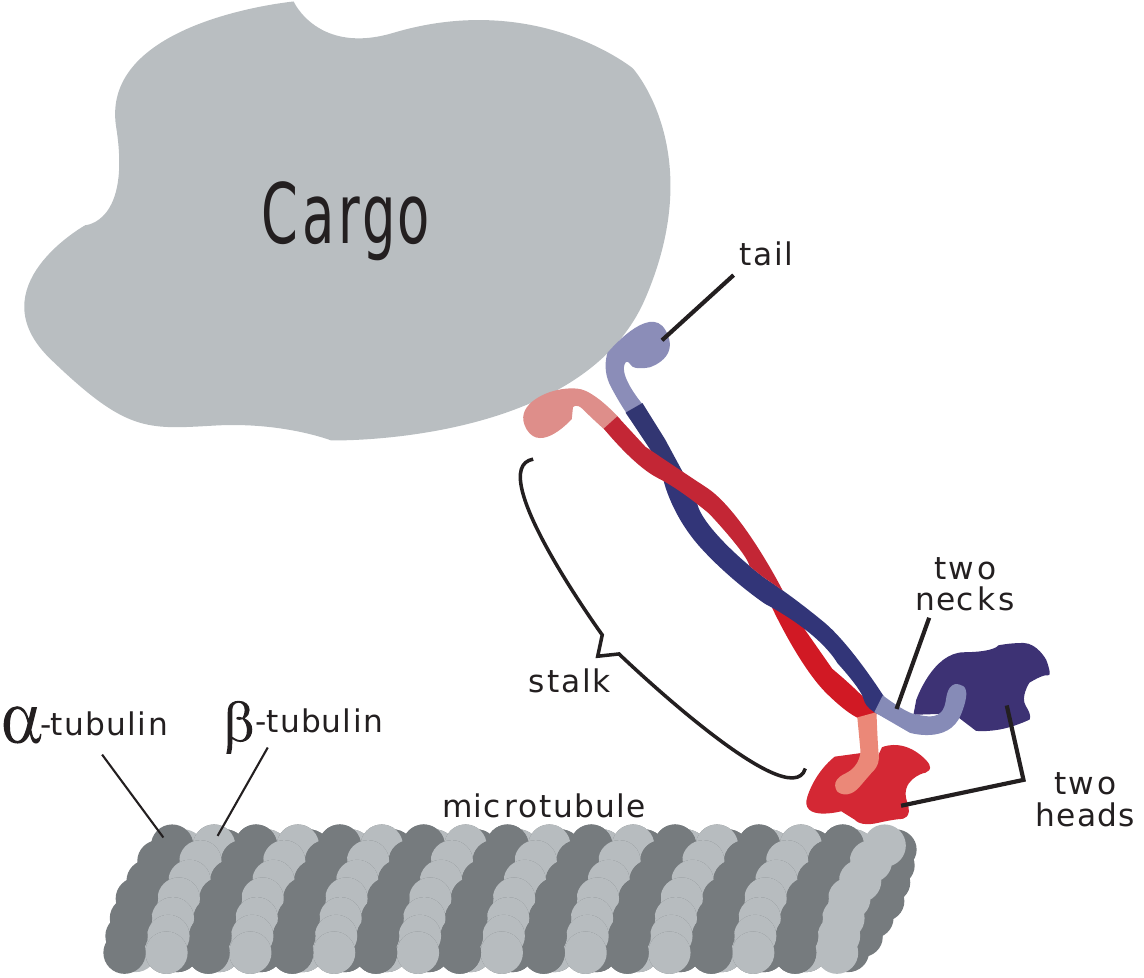}
	\end{center}
	\caption{\label{fig:kinesin} {Structure of the microtubule and the kinesin motor.}}
\end{figure}

One of the examples of the molecular motor proteins is the Kinesin that literally walks over microtubule tracks. Kinesin contains a head domain, a necks domain, a stalk, and a tail as shown in Figure \ref{fig:kinesin}. The head domain binds to the microtubule, the neck domain provides the needed flexibility for the act of walking, the stalk connects the neck to the tail, and the tail domain attaches to a cargo. Hydrolysis of adenosine triphosphate (ATP) in the head domain causes the head domain to walk along the microtubule track in one direction by repeating cycles of conformational changes. ATP hydrolysis detaches the phosphate from an ATP to produce adenosine diphosphate (ADP), which releases energy. This energy is then used by the motor protein to change its conformation and take a single step. During each step one head is detached and moves forward, while the other head stays stationary. Through repeating this cycle, kinesin moves in one direction. In biology, kinesin can carry different cargoes over microtubule tracks from one location in the cell to another location. Therefore, kinesin acts as a locomotive in cells.

If we ignore the detaching probability, the movement of the motor protein can be modelled by one dimensional movement on the microtubule track~\cite{bush-book}. When the molecular motor is in attached state (i.e., on the microtubule) its displacement during time interval $\Delta t$ can be written as
\begin{align}
\ell_{i} = \ell_{i-1} + v_{\text{avg}} \,\Delta t,
\end{align}
where $\ell_i$ denotes  the location of the molecular motor on the microtubule at the $i$th simulation step and $v_{\text{avg}}$ denotes the average velocity. If molecular motors detach from the microtubule they diffuse in the environment and reattach stochastically. Their detaching and reattaching probabilities depend on the properties of the environment, microtubule shape, and curvature.

Inspired by this, it is possible to engineer a microtubule track between a transmitter and a receiver, where kinesin motor proteins would carry information particles from the transmitter to the receiver. For example in \cite{eno11}, it was shown that it is possible to create microtubule tracks \invitro{} in a self-organizing manner, using polymerization and depolymerization. It was also shown that a second approach based on reorganization of microtubules using motor proteins can also be used to create tracks that connect a transmitter to a receiver.

\subsubsection{Microtubule Filament Motility Over Stationary Kinesin}

Although in biological cells kinesin moves over microtubule tracks, in \cite{how89} it was shown that stationary kinesin attached to a substrate can mobilize microtubule filaments. This scheme is very suitable for on-chip MC applications \cite{hiy09}, where the transmitter and receiver are located in a lab-on-a-chip device. For example, recently it was shown that electrical currents can be used to control the speed and direction of the microtubules \cite{duj08,kimE13}. However, unlike kinesin which can naturally carry cargoes, a technique for carrying information particles by microtubule filaments is necessary. 

\begin{figure}[!t]
	\begin{center}
		\includegraphics[width=3.4in] %
		{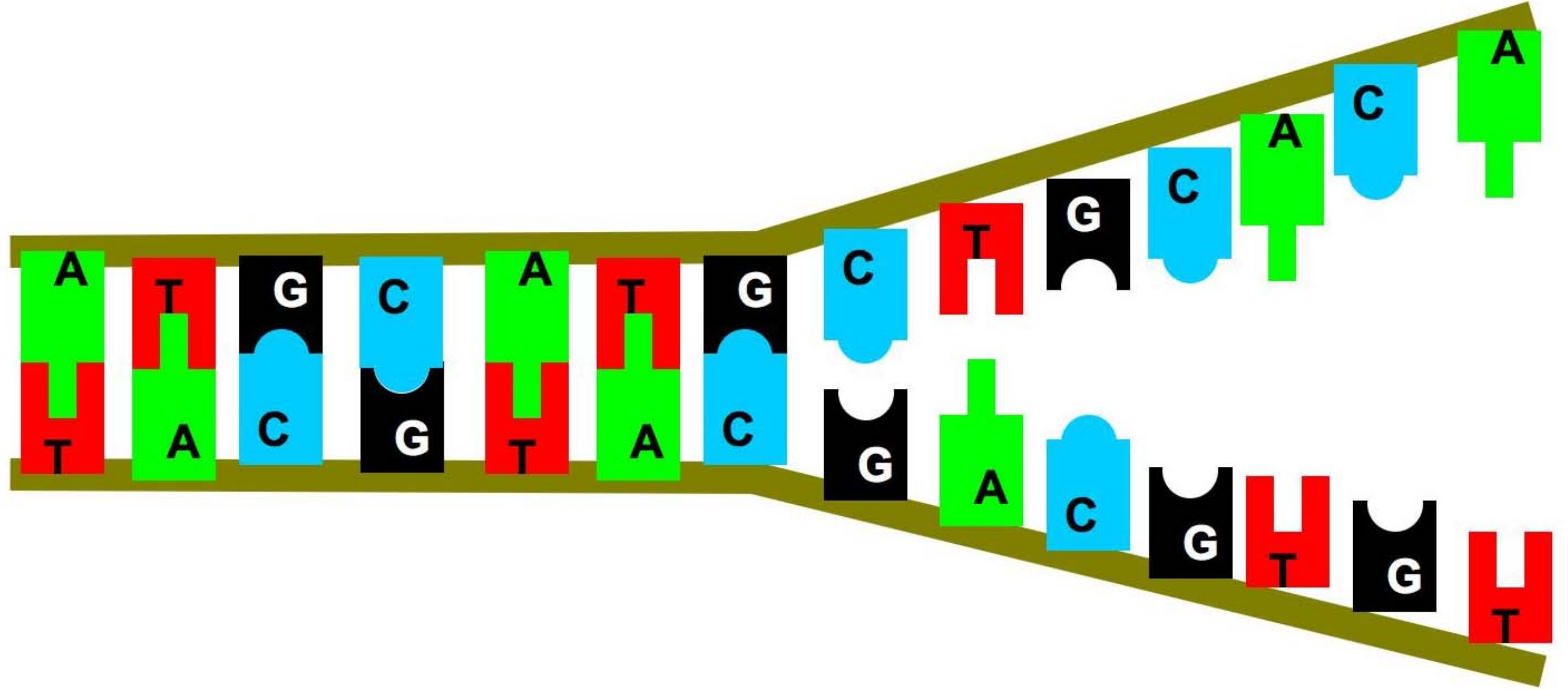}
	\end{center}
	\caption{\label{fig:DNAhyb} {An example of ssDNAs hybridization bonds.}}
\end{figure}

In \cite{hiy09,hiy10LabChip}, DNA hybridization bonds are proposed for carrying cargoes such as vesicles. A single strand of DNA is made up of many nucleotides. Each nucleotide is made up of one of the four possible nucleobase compounds: guanine (G), adenine (A), thymine (T), or cytosine (C). A DNA strand is made up of two single stranded DNAs (ssDNA)s   connected together using hybridization bounds, with adenine bonding only to thymine in two hydrogen bonds, and cytosine bonding only to guanine in three hydrogen bonds. This is called complementary base pairing. The bonding occurs when two ssDNAs with complementary base pairs come close together. Figure \ref{fig:DNAhyb} shows an example of two ssDNAs hybridizing.

\begin{figure*}[!t]
	\begin{center}
		\includegraphics[width=5.8in]{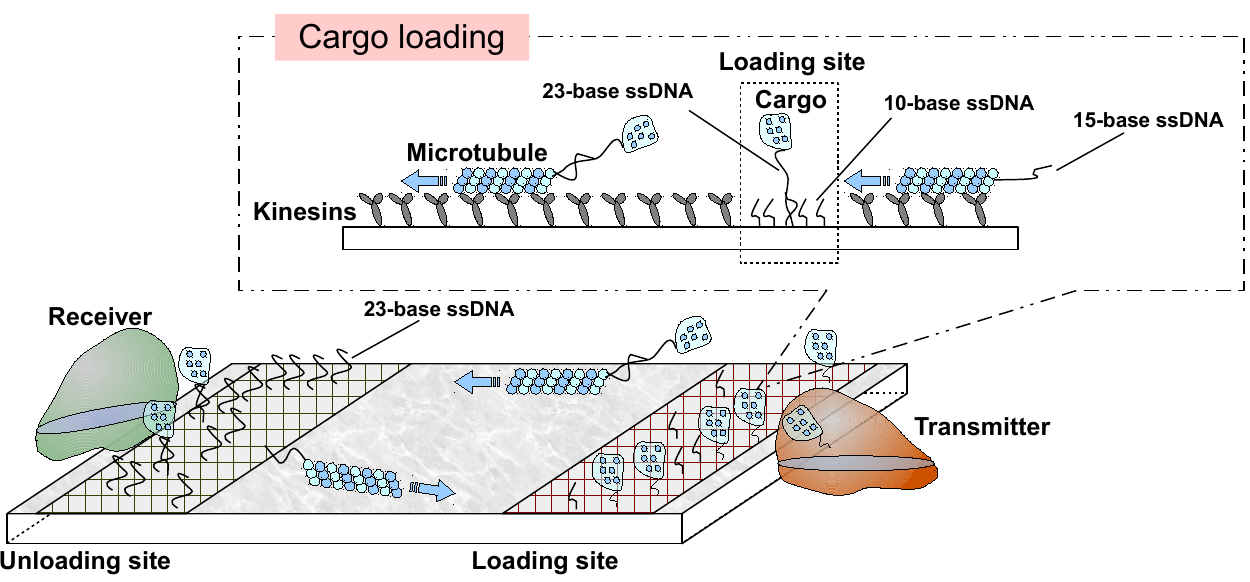}
	\end{center}
	\caption{\label{fig:MTkinDNA} {Cargo transport mechanism for microtubule filaments gliding over stationary kinesin substrate using DNA hybridization bonds.}}
\end{figure*}

In \cite{hiy09,hiy10LabChip}, cargo vesicles, transmission area, receiver area, and microtubules are covered with  ssDNAs. Microtubules moving over a kinesin covered glass substrate are covered with 15 base ssDNAs, and the information particles or the vesicles are also covered with 23 base ssDNAs which are complementary to that of the microtubules' ssDNAs. When the microtubule glides close to an information particle or a vesicle, the two ssDNA sequences hybridize and the microtubule carries the cargo until it gets close to the receiver. The receptor module at the receiver is covered with 23 base ssDNAs, which are complementary to that of the cargo. When a loaded microtubule filament glides close to the receptor module, it will unload the information particle or the vesicle through hybridization bound with the complementary 23 base ssDNA at the receptor module. Since 23 base hybridization bound is stronger than the 15 base hybridization bonds between the cargo and the microtubules, the particles or the vesicles are unloaded at the receiver. Figure \ref{fig:MTkinDNA} depicts the microtubule motility mechanism on a DNA microarray.

In \cite{nit06} and \cite{nit10}, it was shown that the motion of the microtubule filaments over stationary kinesin can be simulated using Monte Carlo simulations. The motion of the microtubules is largely regular, with Brownian motion effects causing random fluctuations. Moreover, the microtubules move only in the $x$ and $y$ directions, and do not move in the $z$ direction (along the height of the channel) because they move right on top of kinesins. During each time interval $\Delta t$, the microtubule filaments travel to new coordinates according to
\begin{align}
\begin{split}
x_{i} &= x_{i\!-\!1} + \stepsize{i} \, \cos \theta_{i} \\
y_{i} &= y_{i\!-\!1} + \stepsize{i} \, \sin \theta_{i} 
\end{split}
\end{align}
where $x_i$, $y_i$, $\stepsize{i}$, and $\theta_{i}$ correspond to the x and y coordinates, step size, and angle of direction of motion at the $\xth{i}$ time interval, respectively. At each step, $\stepsize{i}$ is generated according to a Gaussian distribution with the following parameters
\begin{align}
\stepsize{i} &\sim \mathcal{N}(\vavg \Delta t, \,\, 2D\Delta t),
\end{align}
where $\vavg$ and $D$ correspond to the average velocity of the microtubule and the diffusion coefficient of the microtubule, respectively. The angle $\theta_i$ is also a the sum of the angle in the previous step and the angular change, where the angular change $\Delta \theta_i$ is also a Gaussian random variable and given as 
\begin{align}
\begin{split}
\theta_i &= \theta_{i\!-\!1} + \Delta \theta_i \\
\Delta \theta_i &\sim \mathcal{N}(0, \,\, \vavg\Delta t/\perslen)
\end{split}
\end{align}
where $\perslen$ corresponds to the persistence length of the microtubule's trajectory.  These system parameters are typically ${\vavg = \SI{0.85}{\micro\meter/\second}}$ and ${\perslen = \SI{111}{\micro\meter}}$ as was shown in~\cite{nit06}.

It may also be possible to use actin filaments  and myosin in place of microtubule filaments and kinesin \cite{how01-book}. Myosin is a motor protein that is used for muscle contraction and actin filaments are rod-like proteins. Although they are not exactly similar to kinesin and microtubules, the motion is generated using a very similar manner by employing ATP hydrolysis. In \cite{byu09}, it was shown that silicon nanowires covered with myosin can be used as tracks for actin filaments. Dynein molecular motor and microtubule filaments is another example of motor protein-filament motility \cite{iku14}. It is also possible to use active transport in conjunction with flow based propagation in microfluidic lab-on-chip devices as was shown in \cite{ste14}.

\subsubsection{Bacteria Assisted Propagation}
\begin{figure}[!t]
	\begin{center}
		\includegraphics[width=3.2in]{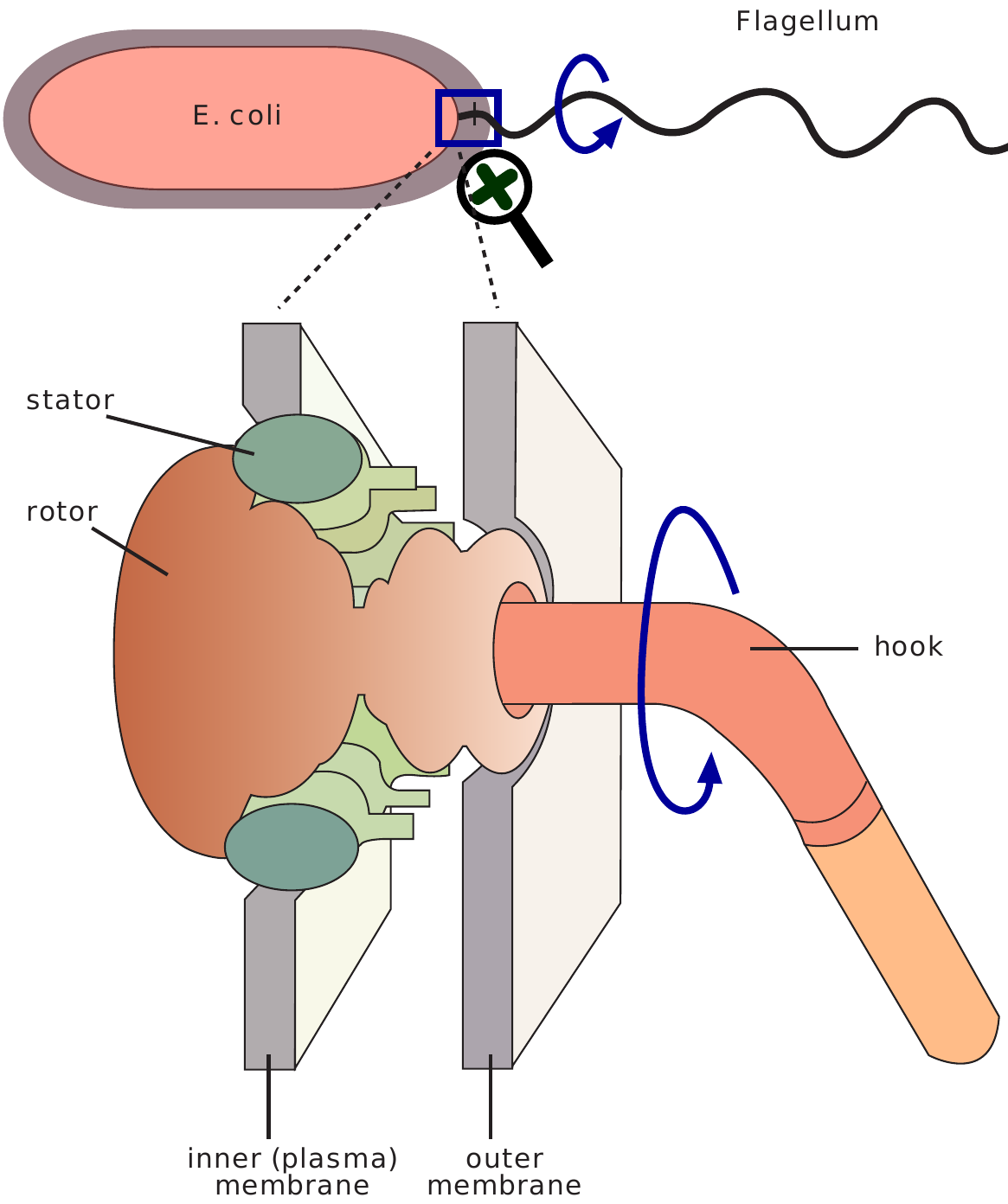}
	\end{center}
	\caption{\label{fig:flagella} {Flagellated bacteria and the structure of flagella.}}
\end{figure}

In \cite{cob10,gre10}, bacteria-based communication was proposed for transferring information particles from a transmitter to a receiver. In this scheme, the information particles are placed inside bacteria at the transmitter and the loaded bacteria are released into the channel. The bacteria then propagate in the channel until they reach the receiver and deliver the information particles. In order to assist this propagation, flagellated bacteria, which are self propelling organisms could be used. Figure \ref{fig:flagella} depicts a flagellated bacteria and the structure of flagella. A Flagellum consists of a filament and a motor section. The motor section rotates the filament, which in turn propels the bacteria in a certain direction. 

\begin{figure}[!t]
	\begin{center}
		\includegraphics[width=0.98\columnwidth,keepaspectratio]{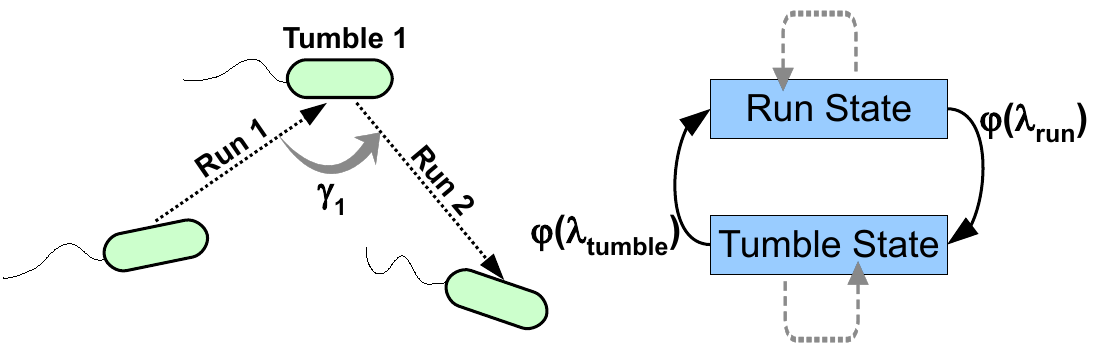}
	\end{center}
	\caption{\label{fig:bacteria_motility} {Bacteria motility dynamics (without attractants).}}
\end{figure}

To guide the bacteria towards the receiver, attractant molecules can be released by the receiver. The bacteria are then attracted towards the receiver by following the concentration gradient and moving towards the increasing concentration of attractant molecules. It is possible to simulate the motion of bacteria as shown in \cite{ber93-book,fry93,gre11}. As shown in the Figure~\ref{fig:bacteria_motility}, the motion of a bacterium consists of two phases: run and tumble states. The displacement is formulated as follows:
\begin{align}
\Delta r_i = v^{\text{run}}_i\,\, T^{\text{run}}_i
\end{align}
where $\Delta r_i$, $v^{\text{run}}_i$, and $T^{\text{run}}_i$ denote the displacement, velocity, run duration for the $i$th run. If there is no attractants in the environment, the run and the tumbling durations (i.e., $T^{\text{run}}_i$ and $T^{\text{tumble}}_i$) are exponentially distributed random variables with means $\lambda_{\text{run}}$ and $\lambda_{\text{tumble}}$, respectively. If there are attractants in the environment, then the mean run length ($\lambda_{\text{run}}$) increases depending on the sensed concentration of the attractant. During the run state, bacteria move nearly straight in the selected direction, but rotational diffusion causes small deviation. This is a Gaussian process with zero mean and $2 \,D_r \,t$ variance, where $D_r$ is the rotational diffusion constant~\cite{ber93-book,wang2008validatingMO}.

Between two consecutive runs, bacteria tumble and choose a new direction. In the tumble state, the new angle for the direction of motion is given by
\begin{align}
\Theta_{n+1} &= \Theta_n + \gamma_n
\end{align}
where $\Theta_n$ and $\gamma_n$ denote the previous angle of motion and the angular change due to tumbling. This change is hypothesized to be uniformly distributed on $[0, \pi]$, but sometimes empirical data is used to obtain a distribution~\cite{ber93-book,wang2008validatingMO}.

\subsubsection{Propagation Through Gap Junction}

Gap junctions are intercellular connections between neighboring cells located at the cell membrane. They are literally gaps that are formed out of two aligned connexon
structures which connect the cytoplasm of two adjacent cells. These gap junctions allow the
free diffusion of selected molecules between two neighboring cells. Their permeability can vary from time to time, allowing different molecules to pass between the cells. Also, they can be closed and reopened by the cell during the lifetime of the cell. 

Intercellular calcium wave (ICW) is one of the main intercellular communication systems observed among biological cells~\cite{alb07}. In this system, after a cell is stimulated, it generates a response to this stimulus with an increased cytosolic Ca$^{2+}$ concentration through the usage of the secondary messenger molecule inositol 1,4,5-triphosphate (IP$_3$). Then Ca$^{2+}$ concentration wave passes to the neighboring cells via usage of IP$_3$ or ATP. There are two main pathways: the internal pathway and the external pathway. In the internal pathway, IP3 molecules diffuse through gap junctions, while in the latter, ATP molecules are released to the extracellular space and diffuse freely to trigger nearby cells. According to experimental studies, it has been shown that these pathways complement each other rather than being alternative approaches~\cite{kang2009spatiotemporalCO}. The frequency and effective range of ICW heavily depends on which pathway is used. Using internal pathway enables frequent fast communication between nearby cells, however reaching distant cells requires the usage of external pathway~\cite{kang2009spatiotemporalCO}. ICW can reach to 200$-$350 $\upmu$m in one dimension (assuming that the cells are aligned linearly) at a speed of 15$-$27 $\upmu$m/s and oscillate in the $10^{-3}-1$ Hz range \cite{scemes2006astrocyteCW, schuster2002modellingOS}. It has been shown experimentally that cells encode the information on the frequency and amplitude components of ICWs~\cite{schuster2002modellingOS}. 

An MC system based on ICW was proposed in~\cite{nak05}. The system is composed of two types of devices, the transmitter-receiver  pair (i.e., source and destination devices), and the ICW-capable intermediary cells. The source device modulates the information onto an ICW wave by generating a stimulus to the closest intermediary cell in the ICW channel. Then the calcium signal propagates through the intermediary cells via the internal and external pathways until the destination device is reached. In~\cite{kur12WCM}, authors presented the capabilities, limitations, and deployment scenarios of calcium signaling system. In \cite{barros2014transmission}, Calcium signaling based MC is proposed and analyzed at different cellular tissues. Authors analyzed the communication system in terms of mutual information where the communicating cells were assumed to reside on a 2-D environment.

\begin{figure*}[t]
	\begin{center}
		\includegraphics[width=1.9\columnwidth,keepaspectratio]{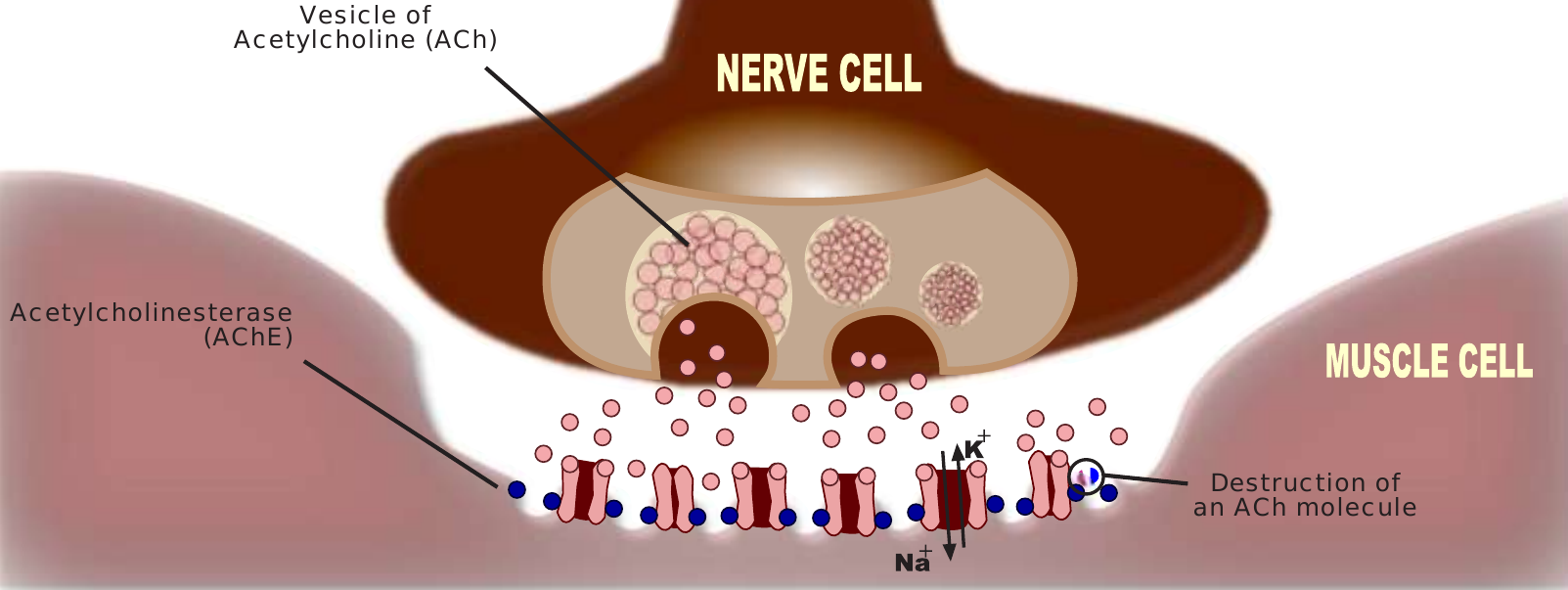}
	\end{center}
	\caption{Representation of neuromuscular junction. Motor neurons release ACh molecules and  they diffuses through the synapse and bind ACh receptors. To keep the communication between the nerve and muscle cell couple, the ACh molecules in the environment are cleaned via AChE molecules.}
    \label{fig:NMJ}
\end{figure*}
\subsubsection{Neurochemical Propagation }

Neurotransmitters are endogenous chemicals that transmit signals across a synapse from a neuron to a target cell. The target cell can be another neuron or another junction cell as in the case of neuromuscular junctions (NMJ). Neurotransmitters are packaged into synaptic vesicles on the presynaptic side of a synapse and then they are released into and diffuse across the synaptic cleft, where they bind to specific receptors in the membrane on target cell \cite{elias2006neuropsychologyCA}.

NMJ, is one of the many occurrences in biological systems where two cells communicate with each other using an intermediary molecule that diffuse in the extracellular environment~\cite{kuran2013aTunnelBA}. NMJ is depicted in Figure~\ref{fig:NMJ} and is a semi-closed environment between a pair of nerve and muscle cell, with a typical length of 10 to 100 nm \cite{freitas-book, keynes2001nerveAM}. When muscles in a specific part of the body need to be contracted, the nerve cells in that region send a signal via neurotransmitters to the muscle tissue through these junctions~\cite{alb07,kuran2013aTunnelBA}. 
\begin{table*}[t] 
\begin{center}
\caption{Comparison matrix of microscale MC propagation schemes.} 
\renewcommand{\arraystretch}{1.2}
\label{tb:micro_propagation_comparison}
\begin{tabular}{p{3.6cm} p{2.7cm} p{1.6cm} p{2.7cm} p{2.7cm}}
\hline
{\bfseries{Propagation Scheme}}& {\bfseries{Based on}} & {\bfseries{Reference}} & {\bfseries{Information Carrier}} &  {\bfseries{Energy Requirement (only Propagation)}} \\  
\hline 
Free Diffusion & Diffusion & \cite{nak08,pie10, pie13} & Molecules &  0\\
Diffusion with First Hitting & Diffusion & \cite{eck07,yilmaz20143dChannelCF} & Molecules &  0\\
Flow Assisted Diffusion & Diffusion $+$ Flow& \cite{sri12} & Molecules &  for Flow\\
Motor Protein over MT & Motor Protein & \cite{eno11} & Vesicle &  1 ATP for $\SI{8}{\nano\metre}$\\
MT over Motor Protein  & Motor Protein & \cite{hiy09,hiy10LabChip} & Vesicle &  1 ATP for $\SI{8}{\nano\metre}$\\
Bacteria Assisted  & Bacteria Motility & \cite{gre10, cob10, lio12} & Bacteria &  for Bacteria Motility \\
Gap Junction  & Diffusion $+$ Gates & \cite{nak05,kur12WCM} & Molecule &  for Triggering Gates\\
Neurochemical  & Diffusion $+$ Enzyme & \cite{freitas-book,kuran2013aTunnelBA} & Molecule &  0\\
\hline
\end{tabular}
\renewcommand{\arraystretch}{1}
\end{center}
\end{table*}

As an action potential reaches the end of a motor neuron, neurotransmitters are released into the synaptic cleft. In vertebrates, motor neurons release acetylcholine (ACh), a small molecular neurotransmitter, which diffuses through the synapse and binds to ACh receptors (AChRs) on the plasma membrane of the muscle fiber. This causes the ion channels at the cell membrane to open, which in turn allows the passage of Na$^+$ and K$^+$ ions. The increased cytosolic concentration of these ions causes the muscle cell to be contracted.  The neurotransmitters stay in the bounded state for some time after which the bond degrades and the ACh molecules are again set free to the NMJ. The degradation of this bond is crucial to the muscle contraction procedure, allowing the muscle to relax and gradually revert back to its original resting position, awaiting further contraction signals. After the degradation of the bonds between ACh and AChR, the neurotransmitter molecules are very likely to re-bond with the receptors. Such an occurrence causes further unwanted muscle contractions and after a few muscle contraction signals, the NMJ will be filled with ACh molecules. This causes the ion channels linked to the AChRs to become inactive, which in turn blocks all further contraction signals. To resolve this issue, the ACh molecules in the environment should be removed from the NMJ after the muscle cell is successfully contracted. This is achieved by using a secondary type of molecule, called acetylcholinesterase (AChE), which resides in the muscle cell. AChE is a special kind of enzyme that is capable of attracting and hydrolyzing the ACh molecules into their two building blocks, Acetate and Choline. Since both of these substructures are incapable of forming bonds with AChR, we can say that in practice AChE molecules remove (or destroy) the ACh molecules from the NMJ.

\subsection{Transmitter/Receiver Mechanisms and Components}
The transmitter and the receiver of microscale MC can be any machine in the microscale to nanoscale dimensions. For example, the transmitter and the receiver can be generated by modifying cells genetically~\cite{you04, bas05, che05, alb97-book}, or by creating artificial cells \cite{dok07, sas10, muk09}. Regardless of the method used, the transmitter requires at least the three components mentioned in the previous section: a unit for generation or storing information particles, a unit for controlling the release of information particles, and a central processing unit. Similarly, the receiver needs at least two components: a detection unit for sensing the information particles, and a processing unit for decoding and deciphering the intended message from the detection unit's measurements. 

Physically, the central processing unit of the transmitter and the receiver may be implemented by synthesizing logic gates and memory into cells as shown in \cite{moo12, siu13}. The information particles can be generated by modifying a metabolic pathway of a biological cell, which then synthesizes and releases specific signaling molecules \cite{bas05, che05}. Transfection, transfer via viral vectors, direct injection to germ line, and transfer via embryonic stem cells are the methods for gene transfer for modifying a metabolic pathway~\cite{cooper2000theCell, huang2013developmentOH, gossler1986transgenesisBM}. Among these methods, viral vectors are a commonly used tool by molecular biologists to deliver genetic material into cells. This process is used for manipulating a living cell to engineer regulatory networks that can be used for communication. For example, in \cite{beal2011automaticCF} a platform for biological system designers  to express desired system functions via high-level programming language was introduced. After forming the genetic regulatory network code, living cells are manipulated via viral vectors. This process can be performed inside a living organism or in cell culture. Viruses efficiently transport their genomes inside the cells they infect for desired function. Main types of viral vectors are retroviruses, lentiviruses, adenoviruses, adeno-associated viruses, and nanoengineered substances \cite{robbins1998viralVF}. 

To control the release timing, a synthetic oscillator can be introduced into a cell \cite{toe10,mon11}, which with the help of the central processing unit acts as the release control module. Therefore, it is possible to have all the components of a transmitter synthesized into cells. Similarly, it is possible to synthesize receptors for a specific type of molecules into cells \cite{hym12,sha07}. Therefore, it is possible to have all the components required for the receiver inside a synthetic cell. More sophisticated processing units can also be designed as shown in \cite{esh09-book}, using novel materials and spin waves (spin waves are propagating disturbances in the ordering of magnetic materials). It is also possible to shrink the current electronic technology to nano scales to create nanoscale processing units~\cite{das07}.

In nature, signals are received via protein structures called receptors. Therefore, these protein structures can be seen as receiver antennas. Receptors are the special protein structures that can bind to specific ligand structures. The binding occurs by intermolecular forces, such as ionic bonds, hydrogen bonds and van der Waals forces. The docking (association) is usually reversible (dissociation). Ligand binding to a receptor alters the receptor's chemical conformation and the tendency of binding is called affinity. The conformational state of a receptor determines its functional state\cite{cuatrecasas1974membraneR}.

Almost all ligand structures in nature capture and remove the information particles from the propagation environment during the detection process~\cite{cuatrecasas1974membraneR}. In some cases the information particles are destroyed after dissociation (e.g., acetylcholinesterase breaks down the molecules in neuromuscular junctions) \cite{cuatrecasas1974membraneR, kuran2013aTunnelBA}.  

Most of the works in receptor theory focus on ligand-gated ion channels and G-protein coupled receptors~\cite{rang2006receptorCP}. Ligand-gated ion channels are a group of transmembrane ion channel proteins which open to allow ions such as Na$^+$, K$^+$, or Ca$^{2+}$ to pass through the membrane in response to the binding of a chemical messenger (i.e., a ligand). These proteins are typically composed of a transmembrane domain which includes the ion pore, and an extracellular domain which includes the ligand binding location (an allosteric binding site). G-protein coupled receptors constitute a large protein family of receptors that sense molecules outside the cell and activate inside signal transduction pathways and, ultimately, cellular responses. 

Cells can also be  created artificially by using liposome vesicles as the membrane encapsulating different functional proteins that together carry the task of the central processing unit, the particle generation and storage unit, receptors, and the particle release control unit for the transmitter and the receiver. Finally, it is possible to develop the transmitter and the receivers by synthesizing novel materials \cite{pat06}.

\subsection{Power Source}

In MC, the transmitter, the receiver, and the propagation process may require power. There are many different techniques for powering these components \cite{mal09}. Sometimes the power is already present in the environment, or can be harvested from the environment through chemical processes. For example, diffusion relies on the thermal energy already present in the channel, and the transmitter and the receiver could be synthetic cells that run on ATP molecules already present in the environment. It is also possible to use other types of chemical reactions to create motion for propagation \cite{sol12}. 

Different components can be powered using an external power source (a power source that is not present or cannot be harvested form the environment). For example, external magnets could be used to power and propel different components \cite{gao10, san11}, while syringe pumps could be used to generate flows that would assist the transport of information particles. In \cite{kur10}, an energy model is derived for MC that is based on diffusion. This model can be used to calculate the power required to generate, encapsulate, and release the information particles at the transmitter. By the help of the proposed model, energy requirement per bit is evaluated with different parameters. Results show that the energy budget affects data rate significantly. It is also shown that selecting appropriate threshold and symbol duration parameters are crucial to the performance of the system.

\subsection{Potential Applications}

There are many potential applications for MC at microscales, such as medical application \cite{freitas-book, mor06}, control and detection of chemical reactions \cite{physChemBook05, dem06}, computational biology \cite{kit02, nob02}, better understanding of biology \cite{mal12}, environmental control and preservation \cite{hir11}, and communication among nanorobots \cite{req03}. The main driving force for engineering MC is the medical field \cite{ata12CM} with applications in lab-on-a-chip devices \cite{cra06}, cell-on-chips devices \cite{ela06}, point-of-care diagnostic chips \cite{yag06}, and targeted drug delivery \cite{vei10}. In many of these applications, communication between different components or devices is the key to unlocking their true potential.

One of the main applications envisioned in medicine is artificial immune system \cite{ata12CM}. In this case, tiny artificial devices are injected into the body, where each device is specialized for a specific task. For example, one device can be specialized to find pathogens, while another is tasked with destroying the pathogens. This is very similar to the immune system, where each immune cell type carries a specific task. Just like the immune system, to function collectively, these devices need to communicate and collaborate with each other. Taking cues from nature, the most promising solution to solving this communication problem would be MC.

Another driving force behind MC is recent advancements in the field of nanotechnology, which is making nanoscale devices such as nanorobots a reality~\cite{req03}.  Limited by their size, nanorobots can perform only simple tasks. Communication and cooperation among nanorobots can result in performing complex tasks~\cite{kos10}. Communicating nanorobots can be used for biomedical engineering applications, where nanorobots inside the body provide significant improvements in diagnosis and treatment of diseases \cite{lea06, cou06, pat06}. In \cite{cav06,mal11}, it is shown that communicating nanorobots can be much more effective at targeted drug delivery than uncommunicative nanorobots. In \cite{cav09},  nanorobots are proposed for detection of brain aneurysm. Nanorobots can also be used for transporting molecular payloads to cells \cite{dou12}, and altering the cell's behavior \cite{elb12}. In~\cite{chahibi2015pharmacokineticMA}, pharmacokinetics of targeted drug delivery systems is analytically modeled through the abstraction of MC. Based on this model, the biodistribution at the target location is estimated with the help of concepts from communication engineering, such as channel delay, path loss, and the drug accumulation in the rest of the body. Also a procedure to optimize the drug injection rate is proposed that is based on the derived models.

Ongoing development of nanoscale electronics, sensors and motors has further advanced the field of nanorobots \cite{cav05}. For example, programmable bacteria can create computation capability for nanodevices \cite{wei01}. Moreover, different techniques can be employed to power nanorobots \cite{mal09}. For example, flagellated magnetotactic bacteria along with magnetic resonance imaging (MRI) can be used as medical nanorobots.  Tracking and controlling of swarms of these bacterias are demonstrated in \cite{mar09}.

\section{Macroscale Molecular Communication}
\label{sec:MacroMC}

Although use of MC for long range communication between nanomachines has been proposed in \cite{gin09}, engineering a truly macroscale MC system has not been considered in the past. In this work, we refer to macroscale MC as a system where the dimensions of the transmitter/receiver, and the distance between them is a few cm or more. These systems have not been considered in the past partly because of the availability of wireless radio technology at these scales, which is very fast and reliable. Nevertheless, there are application for which use of radio technology is not possible or desirable, and MC may be a suitable solution. { {The reliability of the chemical signal in reaching the intended receiver can be greater than radio-based communications in challenging environments where radio signals experience heavy diffraction loss~\cite{guo2015molecularVE}.}
For example, sea water in oceans can be a challenging environment for radio-based communication systems due to the rapid attenuation of the signal~\cite{stojanovic2003acousticUC}.} As another example, use of the radio based sensor networks for infrastructure monitoring in urban areas can be inefficient and unreliable \cite{sta10}. This is because the building infrastructures are typically built from metallic parts and concrete, which can significantly degrade the wireless signals. For example, it has been shown that wireless signals fade very rapidly inside network of metallic pipes that contain many bends \cite{qiu14ICC}. Note that optical wireless communication is not possible for these applications since it requires line-of-sight. However, MC may be used in these applications to transfer information \cite{qiu14ICC}.
In this section, we discuss the physical components required for macroscale MC.

\subsection{Information Particles}
The information particle at macroscales can be any volatile chemical or gas, for over the air applications, or liquids for aqueous environments. For most practical applications, these chemicals must have low toxicity and be safe for humans and the environment. At macroscales, detecting individual molecules would be very difficult. Therefore, in practice concentration of chemicals and chemical mixtures would be used as carries of information.

\subsection{Channel and Propagation}
There are two main forms of propagation at macroscales: diffusion and flow based propagation. In the previous section, we showed that the diffusion and flow based propagation of a single information particle can be simulated using Monte Carlo techniques. At macroscales, instead of a single or few information particles, a large number of particles are used to transfer information. Therefore, a more general formulation, namely the diffusion equations, could be effectively used to model the propagation of chemicals. The diffusion equation allows us to talk about the statistical movements of randomly moving particles. 

The diffusion equation is given by partial differential equation \cite{cran-book} as:
\begin{equation}
\label{eq:diffEq}
	\frac{\partial c}{\partial t} = D \nabla^2 c,
\end{equation}
where $D$ is the diffusion coefficient, and $c$ is the concentration at a particular spatiotemporal location. Therefore, $c$ is a function of $x$, $y$, $z$, and $t$, where the first three variables represent a Cartesian coordinate and $t$ represents time.

This partial differential equation can be solved using different initial conditions. One of the most useful initial conditions corresponds to a point source release of molecules at the transmitter. Assuming $M_0$ molecules are released by the point source suddenly at time $t=0$ 
results in the solution for 1-D, 2-D, and 3-D diffusion as:
\begin{align}
\label{eq:solDEQsourc}
	c(x,t) &= M_0 (4 \pi D t)^{-1/2} \exp \left[ -\frac{x^2}{4Dt} \right], \\
	c(x,y,t) &= M_0 (4 \pi D t)^{-1} \exp \left[ -\frac{x^2+y^2}{4Dt} \right], \\
	c(x,y,z,t) &= M_0 (4 \pi D t)^{-3/2} \exp \left[ -\frac{x^2+y^2+z^2}{4Dt} \right]. 
\end{align}

The diffusion equation can also be solved for other initial conditions \cite{ber93-book}. The channel response can be derived by adjusting the boundary conditions when there is an absorbing receiver~\cite{redner2001guideTF}. The capture function can be obtained by integrating the channel response obtained from the differential equation system with these boundary conditions~\cite{yilmaz20143dChannelCF}. One can find the peak concentration and the peak time, by differentiating the impulse response with respect to time~\cite{lla11b,yilmaz20143dChannelCF}. Moreover, it is possible to find channel inversion in complex s-domain ~\cite{wang2014transmitPS}.

At macroscales, molecular diffusion itself can be a very slow propagation mechanism. For example, the diffusion coefficient of water vapor in air is about 0.3 cm$^2$/s. Therefore, assuming pure diffusion, on average, in 1 second a water molecule propagates about 0.5 cm, in 1 minute 4 cm, and in 1 hour 30 cm. However, it is possible to speed up the propagation by introducing flow.

Other forms of mass transport assist the propagation of particles at macroscales. These include: 
\begin{itemize}
\item {\em Advection} -- Advection refers to transport with the bulk fluid flow \cite{hun03-book}. For example, information particles released inside a duct with air flows, are moved by bulk air flows.
\item {\em Mechanical dispersion} -- Mechanical dispersion, is the result of (a) variations in the flow pathways taken by different fluid parcels that originate in the nearby locations near one another, or (b) variations in the speed at which fluid travels in different regions \cite{jav84-book}. For example, information particles moving through liquid flows in porous materials such as soil follow mechanical dispersion because of different paths they could travel. Similarly, in a pipe where there is a liquid flow, at the boundaries the flow speed could be slower than at the center of the pipe. This effect will disperse the information particles in the pipe.
\item {\em Convection} -- In thermodynamics convection is the fluid flows generated because of difference in temperature \cite{cus09-book}. For example, in a room cold air, which is dense, moves downward while warm air moves upward. It must be noted that convection is also used to specify the combined advection-diffusion process in fluid mechanics. However in this survey, convection is the fluid flows created due to temperature differences.
\item {\em Turbulent flows} -- Turbulent flows are random movements within a fluidic flow \cite{csa73-book}. Because this random movement is similar to molecular diffusion, the motion of particles inside turbulent flows are sometimes called turbulent diffusion.  This is fundamentally different from the processes which determine molecular diffusion: in turbulent diffusion, it is the random motion of the fluid that does the transport, while in molecular diffusion it is the random motion of the information particles themselves. To indicate this difference in causation, the diffusion coefficient for turbulent diffusion is often referred to as the eddy diffusion coefficient. The value of the eddy diffusion coefficient depends on the properties of the fluid flow, rather than on the properties of the information particles. Most important is the flow velocity. Turbulence is only present at flow velocities above a critical level, and the degree of turbulence is correlated with velocity. More precisely, the presence or absence of turbulence depends on the Reynolds Number, a non-dimensional number which depends on velocity, width of the pipe, and the viscosity of the fluid. In addition, the degree of turbulence depends on the material over which the flow is occurring, so that flow over bumpy surfaces will be more turbulent than flow over a smooth surface.
\end{itemize}

When flow is present, the diffusion equation in (\ref{eq:diffEq}) becomes the advection-diffusion equation (also known as convection-diffusion equation or diffusion equation with drift)
\begin{equation}
\label{eq:diffAdvEq}
	\frac{\partial c}{\partial t}+\nabla.(\mathbf{v}c) = D \nabla^2 c,
\end{equation}
where $\mathbf{v}$ is the velocity vector and it is a function of spatial location and time. The $D$ in (\ref{eq:diffAdvEq}) can be the diffusion coefficient for pure molecular diffusion or a combined molecular diffusion and turbulent diffusion coefficients \cite{guh08}. In (\ref{eq:diffEq}) and (\ref{eq:diffAdvEq}), it was assumed that the diffusion coefficient is the same in all spatial dimensions. However, it is also possible to derive analytical solutions when the diffusion coefficient is spatially variable \cite{zop99}. For example, in oceans, the horizontal diffusivity along the water can be $10^7$ times larger than the vertical diffusivity \cite{sak13}.  

The solution for this partial differential equation in (\ref{eq:diffAdvEq}) can be obtained using different initial conditions. For example, assume that the space is 1D, and the flow is constant in the direction of positive $x$ axis only. Moreover, assume that the boundaries are infinite, and $M_0$ information particles are spontaneously released at the origin at time zero. Then the solution of advection-diffusion equation becomes
\begin{equation}
\label{eq:solDiffAdvEq}
	c(x,t) = M_0 (4 \pi D t)^{-1/2} \exp \left[ -\frac{(x-vt)^2}{4Dt} \right],
\end{equation}
where $v$ is the speed in the positive $x$ direction. The advection-diffusion equation can be solved for some other initial conditions as well. 

It is important to note that the diffusion equations and the advection-diffusion equations could be applied at microscales as well, if the number of information particles released by the transmitter is very large. However, in practice the microscale and macroscale environments could be very different. For example, flow at small scales such as microfluidic environments are laminar and behave differently compared to flows at large scale, which could be turbulent. Moreover, the transmitter and the receiver are also different at each scale. Therefore, the end-to-end system responses would be different for macroscale and microscale systems, and each system must be considered separately.

\subsection{Transmitter/Receiver Mechanisms and Components}
At macroscales, the transmitter and the receiver require the same set of components depicted in Figure \ref{fig:MCcomp} at the end of Section~\ref{sec:overview}. For the transmitter, a storage container is required for holding the information particles. It is also possible to generate the information particles using different processes. A mechanism must be set in place for controlled release of chemicals. For example, sprays could be used for controlling the release of information particles. In \cite{col09, mun12}, a technique for releasing complex blends of compounds in specific ratios, which mimics insect pheromones, is developed.

For detection at the receiver, chemical sensors could be used to detect the information particles. For example, metal-oxide gas sensors \cite{boc10} are typically inexpensive sensors, that are capable of detecting the concentration of various types of volatile chemicals and gases. It is also possible to create more sophisticated sensors for detecting mixture of chemicals as shown in \cite{col09Sen}.

The processing unit at macroscales can be a computer or a microcontroller, depending on the application. The power source could be electrical, solar, or any other source. At macroscales, different power sources and processing unit have already been well studied and developed.

\subsection{Potential Applications}
There are some potential applications for macroscale MC. However, because radio communication is a well developed and established technology, most applications must be in areas where radio communication is not possible or not desirable. For example in \cite{sta10}, it is shown that current sensor network technology based on radio signals is not very reliable for some infrastructure monitoring applications. Moreover, there are still numerous problems that must be overcome for radio based communication in underground environments such as mines \cite{for13}. Finally, MC system can be used for sending information in pipes and ducting systems, which could be beneficial in different industries such as oil and gas. 

Macroscale MC can also be used as a tool for studying animals and animal behavior. In nature animals rely on pheromones, which are essentially chemical signals, for simple communication \cite{ago92-book}. For example, ants use chemical trails for navigation and tracking \cite{van98-book}. Inspired by nature, a number of works have used MC to mimic pheromonal communication in insects \cite{col09Sen, col09, ols11, mun12}. Not only this could be a valuable tool to better understand animal behavior, it could also be potentially used to control these behaviors. 

Another potential application for macroscale MC is in robotics, which was initially inspired from pheromonal communication and olfaction from nature \cite{rus98,rus99-book,kuw99}. Generally, previous works in this area can be divided into two main streams: pheromone based communication inspired by nature \cite{kuw99,pur05,pur10,col12}, and plume or chemical tracking robots \cite{rus98,kaz00,ish01,lil01,lar06,wei06,sou08,alb10,alb12}. Besides these applications, MC can also be used for robotics search and rescue operations, and robot communication in harsh environments such as sewer systems.

\section{Communication Engineering Aspect of Molecular Communication}
\label{sec:ComEngMC}

Although MC has been present in nature for billions of years, it was only recently that engineering MC systems has been proposed \cite{hiy05}.  Therefore, compared to modern radio based communication systems, MC is still in its infancy. In this section we review some of the recent work on MC from a communication engineering lens.

In this regard, the MC literature can be categorized based on five partially overlapping topics:
\begin{itemize}
\item \textit{Modulation techniques}: how information is encoded on the information carrying particles and chemical signals.
   \item \textit{Channel models}: Because MC channel can be very different from radio based communication channels, new channel models are required for MC. 
    \item \textit{Coding techniques}: Since the channel is different, there may be a need for new error-correcting codes for MC systems.
   \item \textit{Network architectures and protocols}: The network architecture and protocols used in MC.
   \item \textit{Simulation tools}: Because laboratory experimentation can be time consuming, laborious, and expensive, many different simulation environments have been developed for studying MC. 
\end{itemize}

Over the next sections we survey each category and present the main concepts in each category.

\subsection{Modulation Techniques}

\begin{figure}[!ht]
	\begin{center}
		\includegraphics[width=3.4in]{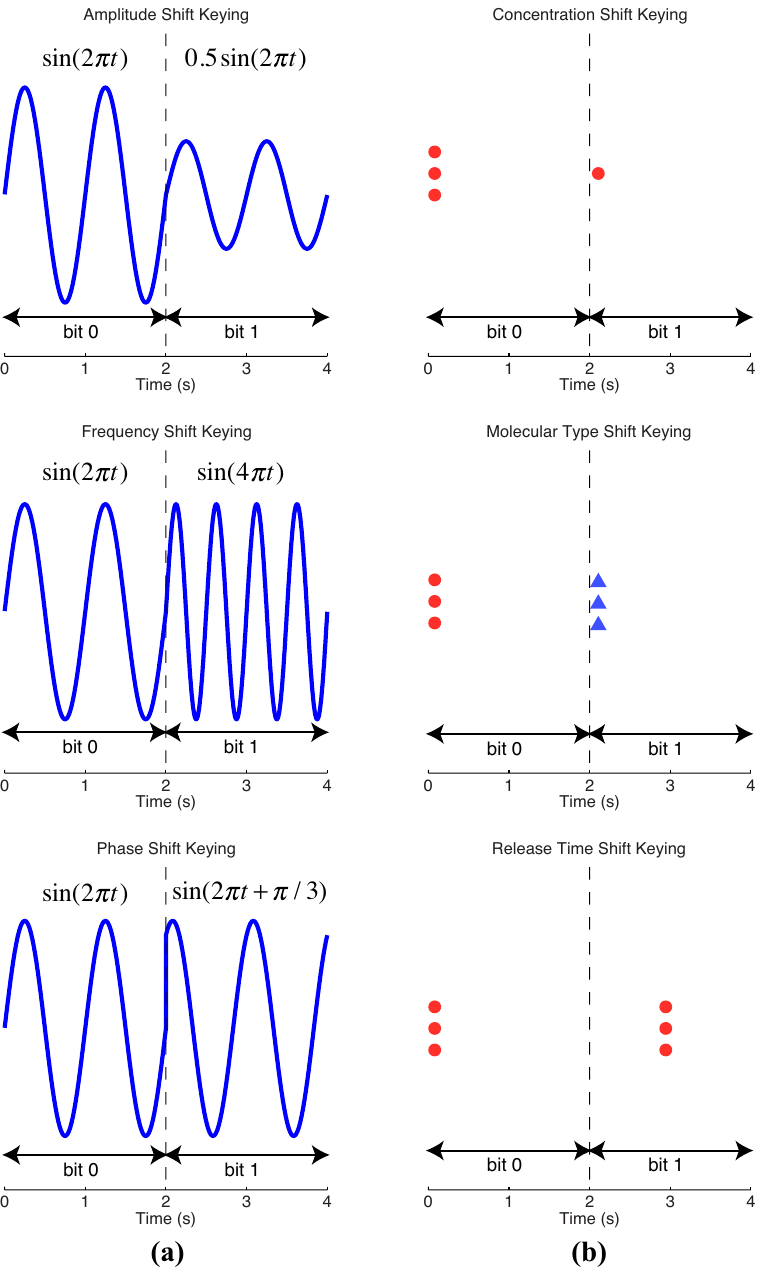}
	\end{center}
	\caption{\label{fig:sinMod} {Modulation techniques in (a) the traditional radio communication (b) molecular communication.}}
\end{figure}

Modulation is the process of varying one or more properties of the carrier signal according to the transmission symbol. In essence, transmission symbols are encoded in changes of the properties of the carrier signal. In traditional radio based wireless communication systems, the carrier of information is electromagnetic waves. Radio waves are sinusoidal signals that can be characterized by their amplitude, frequency, and phase.  The amplitude affects the peak to peak height of the sinusoidal, the frequency affects the number of cycles per seconds, and the phase affects the amount of shift from the origin. Information can be modulated on the amplitude,  the phase, the frequency, or any combination of these parameters. Figure \ref{fig:sinMod}(a) demonstrates the modulation of binary data (0 and 1) using amplitude, frequency, and phase.

\begin{table*}[t] 
\begin{center}
\caption{Comparison matrix of modulation schemes.} 
\renewcommand{\arraystretch}{1.2}
\label{tb:modulation_comparison}
\begin{tabular}{p{4.5cm} p{1.6cm} p{1.7cm} p{3.7cm} p{1.8cm} p{1.9cm}}
\hline
\bfseries{Name} & \bfseries{Abbreviation} & \bfseries{Reference} & \bfseries{Based on} & \bfseries{ISI Reduction} & \bfseries{Molecule Types} \\ 
\hline
On-Off Keying & OOK & \cite{mah10} & Concentration & No & 1\\
Concentration Shift Keying & n-CSK & \cite{kur11, kur12} & Concentration & No & 1\\
Molecular Shift Keying & n-MoSK & \cite{kur11, kur12} & Molecule Type & Moderate & n\\
Isomer-based Ratio Shift Keying & n-IRSK & \cite{kim13} & Molecule Type \& Ratio & Moderate & n\\
Pulse Amplitude Modulation & PAM & \cite{gar11} & Concentration & No & 1\\
Pulse Position Modulation & PPM & \cite{gar11} & Time of Emission & No & 1\\
Emission Time-based Modulation & - & \cite{eck07, hsi13} & Time of Emission & No & 1\\
Time-elapse Communication & TEC & \cite{kri13} & Time of Emission & No & 1\\
Molecular Transition Shift Keying & n-MTSK & \cite{tepekule2014energyEI} & Molecule Type \& Concentration & Yes & 2n\\
Molecular Array-based Communication & n-MARCO & \cite{ata12} & Molecule Order & Yes & n\\
Molecular Concentration Shift Keying & n-MCSK & \cite{arjmandi2013diffusionBN} & Molecule Type \& Concentration & Yes & 2n\\
\hline
\end{tabular}
\renewcommand{\arraystretch}{1}
\end{center}
\end{table*}

In MC, the carriers of information are very tiny information particles (e.g., molecules). In this scheme, information can be modulated on the following properties of the information particles.
\begin{itemize}
    \item  {\em Number of Particles}: Information can be encoded in the number of information particles released. If the number of information particles that are released is very large (e.g., at macroscales), concentration (i.e., the number of particles per unit volume) could be used instead of the number of particles. 
    \item {\em Type/Structure of Particles}: It is possible to encode information in the type (or structure) of information particles released. For example, two different types of molecules could be used to encode one bit of information, or the structure of the DNA could be used to encode a large amount of data. 
    \item {\em Time of Release}: Information can be encoded in the time of release of information particles. Another timing based approach is the pulse position modulation (PPM), which is similar to PPM in optical communication. To generate a pulse, large number of information particles are released almost instantaneously by the transmitter at a specific time.  
\end{itemize}
Figure \ref{fig:sinMod}(b) demonstrates the modulation of binary data (0 and 1) using the number of molecules (concentration), the type of molecules, and their release timing. Just like radio based communication, it is also possible to use any combination of these modulation schemes.   

In the rest of this section we survey some of the recent work on modulation techniques for MC. Generally, these works can be broken down into three main categories: modulation schemes, inter-symbol-interference (ISI) mitigation schemes, and optimal detection at the receiver. Table \ref{tb:modulation_comparison} summarizes the different modulation schemes proposed in the literature.

One of the first works that considered modulation in MC was \cite{mah10}. In this work the authors considered two different modulation schemes for the diffusion-based propagation channels. In the first modulation scheme, a binary bit-0 was represented by concentration zero and the bit-1 by concentration $Q$. This modulation scheme is similar to the on-off-keying from traditional communication. In the second scheme, the concentration of information particles was varied according to a sinusoidal signal with a given frequency, and it is assumed information can be encoded in the amplitude and the frequency of this sinusoidal.

In \cite{kur11} and \cite{kur12}, two new modulation schemes were proposed for diffusion based propagation channels. One was based on the number of molecules, where transmission symbols were encoded in the number of information particles released by the transmitter. This modulation scheme was termed concentration shift-keying (CSK). The second modulation scheme was molecular type-based modulation, where the transmission symbols were encoded in the type of the information particles released. The authors named this modulation scheme molecular shift keying (MoSK), and they proposed hydrofluorocarbon based information particles as an example for this type of modulation. The effects of co-channel inference in a two-sender, two-receiver system was also considered in the work.

The choice of information particles for in-body MC is very important. In \cite{kim13}, the authors proposed aldohexose isomers as an effective information particle for this use. As they pointed out the information-carrying isomers must be selected carefully such that they would not be harmful to the body as the isomers of hydrofluorocarbon proposed in \cite{kur12}. The authors considered isomer based CSK, and isomer based MoSK. They also presented a new modulation scheme, which they called isomer-based ratio shift keying (IRSK). In this scheme, the information was encoded in the ratio of two isomers (i.e., the radio of two information particles).

It is also possible to encode information on the system response pulses of continuous diffusion models. Two different pulse based approaches were presented in \cite{gar11}: pulse amplitude modulation (PAM), and pulse position modulation (PPM). In PAM, a transmission of bit-1 was represented by a pulse at the beginning of the bit interval, and transmission of bit-0 was represented with no pulse. In PPM, a bit interval was divided into two equal halves, and transmission of bits-1 and 0 were represented by a pulse in the first or the second half, respectively. Another work that considered modulation on the system response of continuous diffusion  was \cite{lla13}, where information was encoded in features of the system response. The authors provided expressions for the peak's max, peak's width at half max, and peak's max delay as well as expressions for the energy of the impulse response. The information could then be encoded and detected on these features of the pulses.

Besides using the the number, type and continuous concentration pulse features for encoding information, the release timing of information particles could also be used for encoding information \cite{eck07}. In \cite{hsi13}, the authors considered diffusion based MC, where a hybrid modulation scheme based on type and time of release of particles was proposed. This type of modulation is asynchronous, and could achieve higher information rates than the type based or timing based approaches. To derive their models, the authors presented their modulation scheme as an event-driven system.

Another release timing based modulation scheme called time-elapse communication (TEC) was proposed in \cite{kri13} for very slow networks such as on-chip bacterial communication. In TEC, information was encoded in the time interval between two consecutive pulses (i.e., two consecutive pulses of information particle transmission). The authors showed the feasibility of this technique for communication through bacterial fluorescence. In this setup information particles were released from the transmitter in an on-chip device, where they propagated using microflows until they arrived at the reception chamber. The reception chamber contained bacteria that would produce fluorescence light in proportion to the concentration of information particles. The authors showed that TEC can outperform on-off-keying when techniques such as differential coding was used.

MC channels typically have memory since some of the information particles that are released may remain in the channel and arrive in future time slots. Therefore, another important problem in MC is the ISI due to delayed arrivals of molecules. Hence, modulation techniques should also consider ISI mitigation. One way for minimizing ISI is to combine CSK and MoSK modulation techniques together, which was proposed and named molecular transition shift keying (MTSK) in~\cite{tepekule2014energyEI}. Binary version of this modulation technique utilized two types of information particles $A$ and $B$. The modulator decided which molecule to send depending on the current bit and previously sent transmission bit. Using this scheme, the ISI can be reduced. The same authors proposed pre-equalization technique at the transmitter side for almost eliminating the ISI effect~\cite{tepekule2015novel}. By using two different molecule types,  the proposed technique outperforms the traditional techniques and reduces the bit error rate of the MCvD system significantly.
It is also possible to reduce the ISI using enzymes \cite{noe14}. In this scheme, enzymes that are present in the channel environment slowly degrade the information particles as they move through the environment.  It was shown that this could reduce the ISI and probability of error at the receiver. 

Another ISI-reducing modulation technique based on the order of information particles released by the transmitter was presented in \cite{ata12}. In this scheme, which the authors named Molecular ARray-based COmmunication (MARCO), bit-0 was encoded by releasing information particle $a$ followed by information particle $b$, and bit-1 was encoded by release of $b$ followed by $a$. This reduced the ISI. In a similar work \cite{arjmandi2013diffusionBN}, two types of molecules are used in an alternating fashion to reduce ISI. For example, the transmitter uses type-$a$ molecules in odd time slots and type-$b$ molecules in even time slots. As the molecule types are different in two subsequent time slots, the ISI is significantly reduced.

Demodulation and detection problem at the receiver are an important part of MC systems. In \cite{cho12NanoCom}, a receiver circuit was proposed for frequency shift-keying-modulation presented in \cite{mah10}. For frequency detection, an enzymatic circuit consisting of a number of interconnected chemical reaction cycles was proposed. The circuit was then modeled using reaction-diffusion equations and reaction kinetics equations, which were all partial or ordinary differential equations. The feasibility of the detection scheme was then verified using simulations. 

Optimal detection for MC receivers was considered in a number of works. Optimal receiver design for MoSK modulation scheme was considered in \cite{sha12}. The receiver was optimal in the sense that it minimized the probability of error according to the maximum a-posteriori (MAP) criterion.  The authors first showed that for diffusion based propagation, the channel is linear and time-invariant. They then used this assumption to derive the optimal receiver design. Similarly, in \cite{mah13BSSP,mah13BANC} optimal receiver detection model was derived for binary concentration modulation scheme using MAP criterion. In \cite{kil13}, the authors considered receiver design for on-off-keying modulation scheme based on MAP, and maximum likelihood criterion. They used the continuous diffusion equation and considered the random propagation as noise. Using some simplifying assumptions it was shown that this noise is Gaussian. The sequential detection in the presence of ISI was then achieved by considering the equivalent problem of estimating the state of a finite-state machine. Finally, an optimal detector for the MC channels with enzymes that help reduce the ISI was designed in \cite{noe14Receiver}.

\subsection{Channel Models}
\begin{figure}[!t]
	\begin{center}
		\includegraphics[width=3.4in]{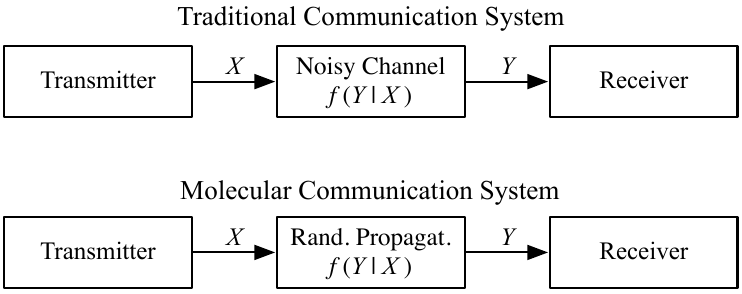}
	\end{center}
	\caption{\label{fig:commChan} {Block diagram representation of communication channels.}}
\end{figure}

Information theory is the mathematical foundation of any communication system~\cite{cover-book}. One of the most basic theoretical models for a communication system is the channel model.  Physically, the channel is the environment over which the transmission signal propagates. Mathematically, the channel relates the transmission information, which would be the channel input, to the received information, which would be the channel output. There is an uncertainty associated with a communication channel. The received signal at the receiver can be different from the transmitted signal because of the noise introduced by the channel. In MC, one source of noise is the random propagation of information particles. Figure \ref{fig:commChan} shows a block diagram representation of the channel.

Unfortunately, most practical MC channels have memory. In this case, channel capacity is calculated differently \cite{gal68}. Let $\mathbf{x}^n$ be a sequence of $n$ consecutive transmission symbols, and $\mathbf{y}^n$ be the corresponding sequence of $n$ consecutive received symbols. If the channel is information stable then the channel capacity is given by \cite{ver94}
\begin{align}
	\mathcal C = \liminf_{n\rightarrow \infty} \sup_{\mathbf{x}^n} I(\mathbf{X}^n;\mathbf{Y}^n).
\end{align}
This makes calculation of channel capacity much more difficult compared to the discrete memoryless channels.

Because channel models play a central role in designing communication systems, one of the first problems that communication engineers tackled was modeling the MC channel \cite{ata07,eck07}. As mentioned in Section~\ref{subsec:MicroProp} different propagation schemes are possible for MC channels. Therefore, for each propagation scheme and modulation technique, a different channel model can be derived. Table \ref{tb:channel_comparison} shows a comparison of most channel models proposed in the literature.

\begin{table*}[t] 
\begin{center}
\caption{Comparison matrix of channel models.} 
\renewcommand{\arraystretch}{1.2}
\label{tb:channel_comparison}
\begin{tabular}{p{5.8cm} p{2.7cm} p{1.6cm} p{3cm} p{1cm} p{1cm}}
\hline
\bfseries{Description and Modulation} & \bfseries{Propagation} & \bfseries{Reference} & \bfseries{Based on} & \bfseries{Considers ISI} & \bfseries{Symbol Set} \\ 
\hline
On-off keying (binary-CSK) & Diffusion & \cite{ata07,ata08,ata09,ata10} & Number of Particles & No & $\{0,1\}$\\
On-off keying (binary-CSK) & Diffusion & \cite{ein2011, ein11ITW, ari11,ata13} & Number of Particles & Yes & $\{0,1\}$\\
On-off keying (binary-CSK) \& particle degradation & Diffusion& \cite{nak12} & Number of Particles & Yes & $\{0,1\}$\\
On-off keying (binary-CSK) \& energy constraints & Diffusion & \cite{kur10} & Number of Particles & Yes & $\{0,1\}$\\
Continuous Diffusion Modulation and Ligand Receptors Demodulation  & Diffusion & \cite{pie10,pie11TSP,pie11TSP2,pie13} & Continuous Diffusion & Yes & $\mathbb{R}^+$\\
CSK using bacterial colonies as Tx and Rx  & Diffusion & \cite{ein12ISIT,ein13TWC,ein13} & Concentration & No & $\mathbb{R}^+$ \\
Contact-Based Neurochemical  & Diffusion & \cite{gun12} & Concentration & No & $\{0,1\}$\\
Reaction Diffusion Channel & Reaction Diffusion & \cite{cho13NanoBio} & Number of Particles & No & $\mathbb{N}$ \\
Channels with spontaneous emission, absorption and particle generation & Diffusion & \cite{mio11} & Continuous Diffusion & No & -\\
Timing Channels - Time of Release Modulation & Diffusion & \cite{eck07} & Hitting Time & No & $\mathbb{R}^+$\\
Timing Channels - Time of Release Modulation & Diffusion with Flow & \cite{sri12,li14} & Hitting Time& No & $\mathbb{R}^+$\\
Time of Release \& Type Modulation & Not Specified & \cite{ros15} & Hitting Time & Yes & $\mathbb{R}^+$ \\
CSK for Diffusion Channels with Flow & Diffusion with Flow & \cite{sha13} & Continuous Diffusion & No & $\mathbb{N}$\\
Kinesin moving over stationary microtubule tracks & Molecular Motors & \cite{moo09b} & Number of Particles & No & $\{0,1\}$\\
Microtubule gliding over stationary kinesin & Molecular Motors & \cite{far11,far12Mona,far14TSP,far12NanoBio} &  Number of Particles & No & $\mathbb{N}$\\
CSK Gap Junction Channels & Gap Junction Diffusion & \cite{nak10,heren2013channelCO,kil13-TNT} &  Number of Particles & No & $\{0,1\}$\\
DNA Propagation Using Bacteria & Bacteria & \cite{gre10, gre11} &  Type/Structure of Particles & No & $\mathbb{N}$\\
Particle Flow Inside Arteries & Diffusion with Flow & \cite{cha13} &  Continuous Diffusion & No & $\mathbb{R}^+$\\
Intercellular Signal Transduction Channels  & - & \cite{eck13} &  Ligand Receptors & Yes & $\mathbb{R}^+$\\
Microfluidic Channel Models  & Diffusion with Flow & \cite{bic13} & Continuous Diffusion  & No & $\mathbb{R}^+$\\
\hline
\end{tabular}
\renewcommand{\arraystretch}{1}
\end{center}
\end{table*}

\subsubsection{Diffusion Channel Models}

In Sections~\ref{sec:MicroMC} and \ref{sec:MacroMC} we described the diffusion propagation scheme for both microscale and macroscale MC. The channel models for diffusion-based MC can be divided into three broad categories: models based on continuous diffusion equations, discrete models, and diffusion models with flow. Besides the propagation, depending on the modulation scheme the channel formulations would be different.

The most popular MC channel in the literature is the time slotted binary-CSK channel employing pure diffusion-based propagation. In this scheme, if the transmission bit is 0, no molecule is transmitted and if the transmission bit is 1, $A$ molecules are transmitted.  In its simplest form $A=1$ (i.e. 1 particle is released to represent bit-1 and 0 to represent bit-0). 

In \cite{ata07}, with using some simplifying assumptions, it was shown that the time-slotted binary-CSK channel can be represented as a binary symmetric channel when ligand-receptors are used for detection at the receiver. The maximizing input probability distribution for this model was presented in \cite{ata08}. The model was then extended to broadcast channels, relay channels
and multiple-access channels in \cite{ata09,ata10}.

Because in diffusion molecules move randomly, there could be ISI between consecutive transmission symbols. In order to reduce the ISI effect a symbol dependent symbol duration could be used~\cite{SIO_NR}, where the transmission time for the current symbol is selected based on the current and the previous symbols. This type of channel was modeled as a Markov chain in \cite{ein2011}, and achievable rates were derived. The authors then completed their analysis in \cite{ein11ITW}, where they modeled the ligand receptor channel as a Markov Chain and derived the capacity and achievable rates for this channel.    

For the binary-CSK channel with $A=1$, if there is no interference with the other molecules from previous transmissions,  the channel could be represented as a z-channel \cite{ari11} . Moreover, the authors in \cite{ari11} considered a very simplified ISI model, where the interference could affect only the next transmission (i.e. the channel memory length is 1). The achievable information rates were then calculated using computer simulations and the results were presented. In \cite{ata13}, a more general ISI model for this channel was considered. It was shown that this communication scheme can be represented as an additive noise channel, where the noise has a Poisson-Binomial probability mass function (PMF)~\cite{Yilmaz_AM}. Using this channel model maximizing input probability distributions was estimated.

In practice, information particles may degrade over time while propagating in the channel. In \cite{nak12}, binary-CSK channels where the molecules degrade over time according to the exponential probability distribution were considered.  Then the channel capacity was compared for different degradation rates and different binary-CSK modulation strategies.  
Another practical problem in MC is the limited power and resources available at the transmitter and the receiver. Energy models for the end-to-end binary-CSK communication channel, which could be used to calculate the maximum number of information particles that can be generated and transported in a particular system was presented in \cite{kur10}. Using this model, an optimal transmission strategy that considered the limited power and resources of the system, and maximized the channel capacity was presented.

Another form of channel model for diffusion-based MC is based on the continuous solution of the diffusion equations presented in Section~\ref{sec:MacroMC}. In this case, the transmission signal is an analog signal $x(t)$, and the received signal is $x(t)$ convolved with the impulse response of diffusion equation plus some noise. One of the first works that considered this approach was \cite{pie10}, where end-to-end gain (attenuation factor) and delay function were derived for continuous molecular diffusion channels with ligand receptors for detection. To derive these terms the authors used techniques from circuit theory to calculate the transfer function for the transmission, diffusion, and reception process. Using these transfer functions they calculated the end-to-end gains and delays.

In practice, solutions for continuous diffusion channels are just the average behaviour across many trials. Moreover, the number of particles that are released by the transmitter are continuous rather than discrete. Therefore,  in \cite{pie11TSP} a more realistic model of continuous diffusion based MC channel was proposed by considering the discrete nature of the number of information particles and randomness of the diffusion propagation. It was assumed that discreteness can be modeled as quantization noise and the random propagation as an additive noise which are added to the solution of continuous diffusion equations.  Similarly, in \cite{pie11TSP2} the authors captured the inherent randomness in ligand-binding reception as noise. Simplifying assumptions and Markov chains were used in the derivation of a closed-form expression of random ligand binding process. 

In \cite{pie13}, the authors derived a closed-form expression for the lower bound on channel capacity of continuous  diffusion-based MC channel in gaseous environments. They showed that this lower bound capacity increases linearly with the bandwidth of the transmission signal. In this derivation, it was assumed that the receiver and the transmitter are perfect, and the derivation was purely based on the diffusion transport process.


Microbial colonies are one of the prominent transmitter and receiver entities in MC. In this scheme, the transmitter and the receiver are essentially made from colonies of genetically modified bacteria that would collectively transmit and receive information particles. The mathematical models for such a receiver were presented in \cite{ein12ISIT}, where the sensing capacity of bacterial nodes was developed. To derive the capacity, it was assumed that the bacteria react to the concentration of information particles by emitting light. The higher was the concentration of the information particles at the receiver node, the higher was the luminance of the bacterial colony. The optimal input distribution and the corresponding sensing capacity were obtained based on the received power. Models for the transmitter and the communication channel between the transmitter and the receiver was considered in \cite{ein13TWC}, where it was assumed that the transmitter was made of colony of bacteria and the transmission of information particles was a two stage process. It was assumed that signalling molecules (type A molecules) were used to control the number of information particles (type B molecules) released by the transmitting colony. The released information particles then diffused until they arrived at the receiver colony, where they were detected by ligand receptors, which then initiated a fluorescent light generation response that would be dependent on the concentration of the received molecules. The end-to-end channel model and some expressions for the capacity were then presented. 


Another work that considered the transmitter and the receiver entities in diffusion-based MC is \cite{gun12}, where the transmitter and the receiver were assumed to be genetically modified cells that are moving themselves. The information was transported through a three phase process: collision, adhesion, and neurospike transmission. In this scheme, two cells (the transmitter and the receiver) moved according to the laws of diffusion, until they collided and stuck to one another. The information was then transported from one cell to another using molecular neurospike communication. Through some simplifying assumptions, the authors derived the channel models for this communication scheme.


In practice, information particles may chemically react with other molecules present in the channel. In \cite{cho13NanoBio}, a model based on reaction-diffusion master equation, which is a well known model in physics and chemistry for jointly modeling diffusion and chemical reactions, was proposed. The model which is called reaction-diffusion master equation with exogenous input (RDMEX), discretizes time and space and uses a Markov model, where the states of the model are the number of information particles in discrete locations of space. During each discrete time step, the state of the Markov model changes according to the laws of diffusion and chemical reactions. 
Another work that considers the effects of the channel on the information particles is \cite{mio11}, where a stochastic model for three different environmental effects are considered: particle absorption (where the particles are absorbed in the environment), particle generation (where a single information particle becomes two information particles through reactions in the medium), and spontaneous emission (where information particles are spontaneously generated through chemical reactions). Through simplifying assumptions, a model was derived using birth-and-death processes. The model was then applied to diffusion based MC with and without flow.

As mentioned in the previous section, it is possible to modulate information on the time of release of information particles. One of the first works that considered diffusion-based timing channel was \cite{eck07}, which was based on the 1-dimensional first hitting time distribution. The channel capacity for this model was derived and it was shown that for this simplified channel, more than one bit of information can be transmitted per information particle. As mentioned in Section~\ref{subsec:MicroProp}, flow could be used to increase the speed of propagation. 
In \cite{sri12}, it was shown that such a channel, employing time of release modulation, could be modeled with an additive inverse Gaussian noise (AIGN) term. Lower and upper bounds for the channel capacity of AIGN channel was then derived. It was shown that these bounds were tight for low flow rates, but the bounds became separated as the flow rate increased or as it approached zero. The derived channel model was then used to design a receiver based on a maximum likelihood estimator. Tighter bounds for AIGN channel were presented in \cite{li14}. A general formulation for timing and type based channels, where information is encoded on both time of release and type of particles is presented in \cite{ros15}.

\subsubsection{Other Channel Models} 

Many different channel models are proposed for diffusion based MC. However, for other propagation schemes presented in Section~\ref{subsec:MicroProp}, there are not many channel models in the literature. In this sections, we survey some of the works on these propagation schemes.

One of the most promising propagation schemes besides diffusion is active transport using molecular motors. Kinesin-microtubule motility is the most widely used form of active transport. As it was discussed in Section \ref{sec:MicroMC}, two forms of transport are possible: kinesin moving on stationary microtubule tracks, or microtubule filaments gliding over kinesin-covered substrate. Generally, active transport is a suitable propagation scheme for on-chip applications.
   
First, we consider channel models for kinesin moving over microtubule tracks. One of the early works on modelling this channel is \cite{moo09b}, where simulations were used to compare the arrival probabilities of diffusion channels, and active transport channels based on a microtubule track that connects the transmitter to the receiver. In this setup, kinesin motor proteins carry the information particles on the microtubule tracks. A hybrid approach based on combination of these propagation schemes was also considered. It was shown that diffusion had low rate of delivery and lower channel capacity for separation distances in the order of micrometers compared to the active transport and the hybrid schemes.

Second, we consider models for microtuble filaments moving over kinesin covered substrate. In \cite{far11}, a simple mathematical model was presented for this channel. In particular, these models were developed to decrease the simulation time of these channels. which were used in previous works. A new Markov chain channel model  was presented in \cite{far14TSP}, where the results of \cite{far12Mona} were improved. It was shown that the Markov chain channel model can reduce the simulation time, and can be used to calculate different channel parameters such as channel capacity fairly accurately. In \cite{far12NanoBio}, on-chip MC channels were considered in confined environments. In particular, the channel capacity of diffusion based, flow based, and kinesin-microtubule based MC channels were compared using Monte Carlo simulations. It was shown that at short distances, diffusion and flow based propagations achieve higher channel capacities, while at longer distances active transport and flow based propagations achieve higher channel capacities.

Gap junction channel is another propagation scheme that has been considered in a number of previous works. In this scheme, it is assumed that the transmitter and the receiver are cells or artificial lattice structures that are connected together through a series of similar cells or artificial lattice structures. Each cell or lattice structure is connected to its neighbors through gap junction channels. Calcium ions are generated by the transmitter and propagate through the gap junction channels, traveling from lattice to lattice until they arrive at the receiver. Binary-CSK modulation scheme in such a MC system was considered in \cite{nak10}. The channel was then simulated and achievable information rates were presented. In \cite{heren2013channelCO}, the channel capacity of the calcium signaling system based on an inter-cellular calcium wave model for astrocytes was investigated. Calcium waves are formed by the cytosolic oscillations of Ca$^{+2}$ ion concentration, which propagates through neighboring cells via secondary messenger molecules.  Channel capacity of the investigated system was analyzed under different noise levels and symbol durations. In \cite{kil13-TNT}, the gap junction communication between cardiac muscle cells, called cardiomyocytes, was modeled. In particular the propagation of action potential, which helps to regulate the heart beat was modeled as a communication channel and then simulated to calculate the channel capacity. 

Bacteria are another effective tool for actively transporting information particles from the transmitter to the receiver. An MC channel in which information particles such as DNA strands are embedded in flagellated bacteria that then transport the particles was modeled in \cite{gre10, gre11}. It was assumed that the receiver uses attractant molecules to attract these bacteria. The motion of the bacteria from the transmitter to the receiver was then modeled and simulated.

Communication theoretic channel modelling has been used in a number of works to represent interesting biological processes.  In \cite{cha13}, a channel model for blood flow through arteries was developed using harmonic transfer matrix and some simplifying assumptions. Two different models were used for small and large arteries, and the models were then combined to form a complete model of blood flow through the arteries in the body. A model for drug injection propagation in the arteries was then presented based on this model. In \cite{eck13}, simple intercellular signal transduction channels were considered, where information particles were detected using ligand receptors. The paper abstracted the concept of propagation and assumed the input to the channel is the concentration level at the vicinity of  the receiver, which was equipped with ligand receptors. The detection process was then represented as a discrete-time Markov model. It was shown how the capacity can be obtained for this channel. It was also shown that the capacity achieving input distribution is iid, which is unusual for channels with memory. Furthermore, it was demonstrated that feedback does not increase the capacity of this channel.   

Microfluidics is another important area where MC can have a high impact. Channel models for flow-based MC in microfluidic channels was presented in \cite{bic13}. First, the transfer function of the straight channel, the turning channel, bifurcation channel, and  combining channel was derived. Then it was shown that the overall transfer function of any channel configuration (i.e. any combination of straight, turning, bifurcation, and combining channels) can be calculated using the transfer function of these channels. Using this model, the authors demonstrated that finite impulse response (FIR) filters could be implemented on a microfluidic device.

\subsection{Error Correction Codes}

In traditional communication systems, channel codes are used to mitigate the effects of noise and fading that is introduced into the system by the channel and electronic components. In essence, channel coding introduces redundancy which can then be used to detect and correct errors. For example, in a simple repetition code, a binary bit-0 is encoded as 000, and a binary bit-1 is encoded as 111. At the receiver the majority number of bits are used to decode the bit. For example, 011 is decoded as bit-1, and 010 is decoded as bit-0. It is fairly trivial that such a simple repetition code can correct one bit errors.
\begin{table*}[t] 
\begin{center}
\caption{{Comparison matrix of coding schemes for MC.}} 
\renewcommand{\arraystretch}{1.2}
\label{tb:coding_comparison}
\begin{tabular}{p{3.1cm} p{2.9cm} p{1.6cm} p{2.7cm} p{3.0cm}}
\hline
{\bfseries{Coding Scheme}}& {\bfseries{Based on}} & {\bfseries{Reference}} & {\bfseries{Energy Considered}} &  {\bfseries{Error Types}} \\  
\hline 
Hamming & Hamming Codes& \cite{lee12} & Yes &  Additive Noise\\
MEC (Hamming) & Hamming Codes& \cite{bai2014minimumEC} & Yes &  Additive Noise\\
MC-CE & Convolution Codes & \cite{mah13} & No &  Additive Noise\\
MoCo & MoCo Distance & \cite{ko2012newPF} & No &  Intra- and Inter-codeword\\
ISI-free Code  & One to Many Codewords & \cite{shi13} & No &  Intra- and Inter-codeword\\
Zero Error Code  & Discrete Time Queues & \cite{kov14} & No &   Intra- and Inter-codeword\\
\hline
\end{tabular}
\renewcommand{\arraystretch}{1}
\end{center}
\end{table*}

{\em Capacity-approaching codes} are codes that allow information transmission rates that are very close to the maximum theoretical limits of the channel (i.e., channel capacity). Many different capacity-approaching codes such as low density parity check (LDPC) codes and turbo codes have been developed for the traditional communication systems \cite{moo05}. However, the encoder and the decoder for these channel codes are computationally complex, and as such they may not be practical for many MC systems, specially at microscales. Furthermore, because the nature of the noise is different in MC channels, it is not clear if better error correcting codes exist for these channels. For example, MC channels typically have memory. Only a limited amount of work has been done on channel codes for MC, and there are still many open problems in this area.

The early works on coding schemes for MC considered applying the previously devised coding schemes for radio communications. One of the early works that considered channel codes for MC systems was \cite{lee12}, where Hamming codes were proposed as simple error correcting codes for on-off-keying in diffusion based MC. First, it was shown that these error correcting codes can improve the probability of bit error rate when a large number of information particles were used for representing each bit. Uncoded transmission outperformed coded transmission, however, when smaller number of molecules were used to represent each bit. This occurs because of the extra ISI introduced by the extra parity bits. The authors then modeled the extra energy needed for transmitting the extra parity bits and showed that coding was not energy-efficient when the separation distance between the transmitter and the receiver was small. An energy efficient version of this coding scheme was presented in~\cite{bai2014minimumEC}. The proposed method minimized the average code weights (i.e., more ${\mbox{bit-0}}$ is utilized). According to the results in~\cite{bai2014minimumEC}, energy consumption was reduced with a cost of larger codeword lengths. In \cite{mah13}, convolutional coding schemes were applied to the same channel and it was shown that convolutional codes could improve the bit error rate of MC channels.

\begin{figure}[!t]
	\begin{center}
		\includegraphics[width=0.90\columnwidth,keepaspectratio]{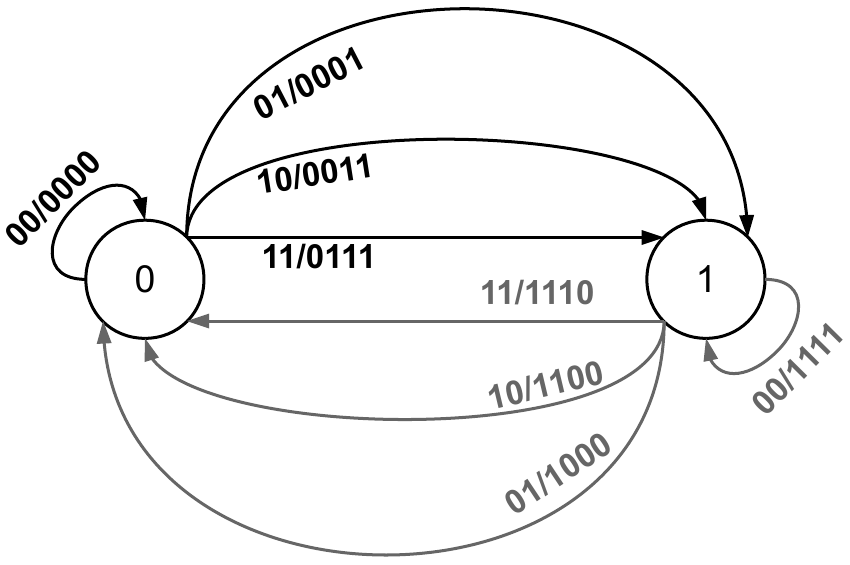}
	\end{center}
	\caption{\label{fig:isiFreeCoding} Encoder of the example ISI-free (4,2,1) code. For example, $01\,\,10\,\,11$ message is encoded as $0001\,\, 1100 \,\, 0111$. First block ($01$) is encoded as $0001$ with a transition to state 1. Therefore, second block ($10$) is encoded as $1100$. Hence, you reduce the deteriorating effect of the ISI by considering the final bit of the codewords.}
\end{figure}
A few works have developed new coding schemes tailored for MC channels. In \cite{ko2012newPF}, molecular coding (MoCo) distance function was proposed. MoCo considers the transition probabilities of the codewords that is affected by the random walk of the molecules. The MoCo code construction aims to find the codewords that maximize the minimum pairwise MoCo distance. In \cite{shi13}, a new code family called ISI-free code family was introduced for diffusion based MC channels with drift. It was assumed that the type of the information particles is used to encode information. In this coding scheme, the crossovers up to level-$\ell$ between consecutive codewords and within the current codeword were eliminated. This coding scheme was defined with $(n,k,\ell)$, where $n$, $k$ and $\ell$ denote the block length, message length, and correctable crossover level, respectively. An example encoder for ISI-free (4,2,1) coding scheme is given in Figure~~\ref{fig:isiFreeCoding} and it utilizes the last bit of the previous codeword. Having two different codewords for each message suppresses the inter-codeword interference since the last $\ell$ bits of the first codeword are identical to starting $\ell$ bits of the following codeword. It was shown that these code families can improve the bit-error rate performance of MC channels compared to uncoded transmission, under similar throughputs. In~\cite{lu2015comparisonOC}, the authors compared hamming, LDPC, and cyclic codes for diffusion channel. Another work that developed a new coding scheme for MC timing channels was \cite{kov14}. In this work, zero-error codes were developed which guaranteed successful recovery of information bits at the receiver, assuming the channel has finite memory.

\subsection{Architectures, Protocols, and Optimal Design}

Because MC is still in its infancy, most of the research has been focused on modeling the channel as well as the modulation and coding schemes necessary for setting up a reliable communication link. However, once the link is established other important aspects of communication networks are the network architecture and protocols. The Open Systems Interconnection (OSI) model is a conceptual model that characterizes and standardizes the internal functions of a communication system by partitioning it into abstraction layers. For example, TCP/IP model of the Internet partitions the communication system into 4 layers: link layer, Internet layer, transport layer, and the application layer.

When we are considering MC, one might think that we may be able to use the same layered architecture over an MC physical link. However, one issue is that we may not be trying to connect electronic devices in a similar fashion as we do in radio telecommunication systems. Instead, we may be interconnecting biological entities, where each device has different limitations and requirements. Moreover, the communication link itself as well as the physical and chemical properties of environment could ultimately impose limitation on how the system could be abstracted into different layers. Therefore, it is not perfectly clear if using the same layering structure would be appropriate. Even if we use the conventional layering architectures, it is not clear how each layer should communicate with the layer above or below. A detailed survey on how MC networks could be layered was presented in \cite{nak14TNB}. 

To solve some of these issues, IEEE Standards Association has started the IEEE P1906.1 project to provide recommended practice for nanoscale and MC framework \cite{IEEEP1906.1}. Some of the goals of this project is as follows. Provide a common framework that would greatly aid in developing useful simulators for MC. This includes interconnecting systems of multiple types of simulators. Provide a common abstract model that would enable theoretical progress to proceed from different disciplines with a common language. This framework serves as a recommended practice for additional nanoscale networking standards as industry becomes more involved. In the rest of this section, we highlight some of the important works on network architecture, protocols, and optimal design.

\subsubsection{Network Architecture}
One of the early works that looked at network architectures for MC was \cite{wal09}. The authors assumed that the intended transmission information can be divided into two categories: sensor data, and command data. They then assumed that the command data has higher priority and as such would be encoded with error correction codes, where as sensor data are uncoded. To solve the computing problem required for carrying out the network protocol, the authors proposed DNA and Enzyme based computations. In particular, DNA based computations was proposed for application interface stack, which encodes messages and addresses, network stack which routes messages, and error correction stack which encodes and decodes error correction codes. Enzyme based computation was proposed for link switching stack.

Another work that explored the network architecture for MC was \cite{gre10}, where each nano-node could communication with other nano-nodes over the short range (nm to $\mu$m) using diffusion, the medium range ($\mu$m to mm) using bacteria transport and catalytic nanomotors, and the long range (mm to m) using pheromones. In the proposed architecture, it was assumed that each nanodevice was connected to a gateway over the short range and the gateways are distributed over the medium to long range. It was assumed that instead of bits, DNA base pairs are used for addressing and carrying information.

Genetically engineered bacteria can play an important role in MC systems. A number of works explore network architectures based on genetically modified bacteria. In \cite{lio12}, the authors proposed a routing mechanism for bacterial based MC. In this scheme, it was assumed that each node in the network released chemoattractant to attract the bacteria from neighboring nodes. It was also assumed that information particles are DNA sequences that are embedded in bacteria. Node A embedded a DNA sequence inside bacteria that pass through neighboring node (node B) until they arrive at the destination (node C). At the relay node (node B), it was assumed that the bacteria transfer their information to a new set of bacteria that are sensitive to node B's neighboring chemoattractants. The DNA sequences were transferred from bacteria to bacteria using bacterial conjugation, a process that involves transfer of single stranded DNA from one cell to another. Since there may be errors during conjugation, in \cite{bal13} use of antibiotics were proposed to kill bacteria with incomplete DNA sequences. In this scheme bacteria resistance genes were inserted at the end in the DNA sequence carrying the information. If bacterial conjugation results in an incomplete DNA sequence, the bacteria carrying the information particles was assumed to be killed by the antibiotics. 

\subsubsection{Multiuser Environments}
In a system with multiple sources and multiple receivers, it is important for the bacteria to deliver the DNA sequence to the correct destination. In \cite{moo11}, the authors proposed establishing a coordinate system using beacon nodes. This idea was analogous to the way global positioning system (GPS) works. It was assumed that each beacon node releases a specific type of chemical at a specific rate. Therefore, there is a concentration gradient generated by each beacon. Distances from a receiver to a beacon was measured using this concentration gradient and the fading of concentration over a distance. The information particles (DNA sequences) was then assumed to be carried by genetically engineered bacteria to the correct destination using chemotaxis.  

Another work that considers networks with multiple transmitters and receivers was \cite{nak13}, where the authors considered optimizing the transmission rate of each transmitter such that the overall throughput and efficiency was maximized. In their model it was assumed that there are $N$ transmitters and $M$ receivers. At the receiver, it was assumed that the information particles were detected through Michaelis Menten enzyme kinetics. The throughput was defined as the number of information particles processes at the receiver per unit time. The efficiency was defined as the throughput divided by number of information particles transmitted per unit time. A simplified optimization model where all transmitters are at origin and all receivers are at distance $r$ was presented, and the solution was provided with an upper bound on throughput and efficiency. A feedback mechanism was also presented that would adjust the optimal transmission rate in dynamic environments.

\subsubsection{On-chip MC}
Another important area of interest in MC is on-chip application. Generally, works in this field can be divided into two categories: those based on active transport propagation schemes, and those based on microfluidics. In \cite{eno11}, an architecture for design of self-organizing microtubule tracks for on-chip MC was proposed. In this scheme, microtubule tracks act as railways or wires connecting the transmitter to the receiver. In particular, two different approaches were considered: one based on polymerization and depolymerization of microtubules, and the other based on molecular motors for reorganizing the tracks. Preliminary \invitro{} experiments were conducted then to investigate the feasibility of the proposed techniques.

For on-chip applications, it is also possible to employ stationary kinesin and microtubule filaments for information particle transport. In \cite{far11NanoCom}, the authors found an optimal design for the shape of the transmission zone and the optimal size for encapsulated information particles. All these parameters were optimized through maximizing the channel capacity of the communication system. In \cite{far11Bionet,far12Mona}, it was shown for rectangular channels, the square-shaped channel maximizes channel capacity. In \cite{far12NANO}, the results were extended to include regular polygon shapes, and it was shown that circular-shaped channel tends to be capacity maximizing for these systems. Finally, the exact optimal channel dimensions based on system parameters can be calculated using the model presented in \cite{far15TNANO}.  

Droplet microfluidics is another promising form of communication for on-chip applications. One of the first works that considered the problem of networking in droplet microfluidics was \cite{leo12}. The authors first gave an overview of this field, and then discussed some of the issues and problems in creating network architecture for hydrodynamic based droplet microfluidics. It was shown that a ring topology, where each processing module is connected to a main ring through a microfluidic network interface (MNI) circuit could be an effective architecture for networking different components in these systems. In this scheme, the main ring starts and ends at the central processing unit, and the network interface circuits control the flow of droplets through and from each processing module. A general introduction to hydrodynamic microfluidics was also provided in \cite{bir13}.

The similarity between Ohm's law and Hagen-Poiseuille's law was used in \cite{don13} to design a switch circuit for the MNI. In this scheme, the circuit was designed in such a way that depending on the distance between a header droplet and a payload droplet, the payload droplet was either delivered to the corresponding processing unit or it continued flowing in the ring. A medium access control circuit was also proposed for releasing a payload into the ring such that the released droplet would not collide with the other droplets flowing in the ring. In \cite{leo13}, channel capacity expressions for the different droplet encoding schemes were provided.

For many practical MC systems such as on-chip molecular communication, the number of information particles may be limited. In \cite{dem13}, MC in confined environments was considered, where it was assumed that the number of information particles within the environment was constant. Therefore, different nodes needed to harvest the information particles that were within the environment for communication. The authors then simulate this particular architecture and present some preliminary results based on different harvesting and communication protocols.

\subsubsection{Protocols}
It is possible to exploit the properties of MC channels to develop clever protocols. In \cite{moo12-TSP}, a protocol for measuring distance between two nanomachines by exchanging information particles and measuring their concentration was proposed. In this scheme, first node $A$ released $N_a$ particles of type-$a$ into the environment, where they diffused. At node $B$ the change in concentration of type-$a$ particles was measured. Then, $N_b$ particles of type-$b$ were released into the environment according to some property of the measured concentration of ${\mbox{type-}a}$ particles. By measuring the concentration of type-$b$ particles back at node $A$, the distance was estimated. Different detection processes such as round trip time and signal attenuation were considered.

Another interesting protocol for blind synchronization between nodes in a diffusion-based MC system was proposed in \cite{sha13CL}. In particular, an MoSK channel was assumed, where during each symbol duration different molecules are used to encode different messages. Then a non-decision directed maximum likelihood criterion was used to estimate the channel delay, which would be used for synchronization. The Cramer-Rao lower bound for the estimation was also derived, and the results were compared with simulations.

\subsection{Simulation Tools}
\begin{table*}[t] 
\begin{center}
\caption{Comparison matrix of MC simulators.} 
\renewcommand{\arraystretch}{1.2}
\label{tb:sim_comparison}
\begin{tabular}{p{3.7cm} p{1.95cm} p{1.95cm} p{1.95cm} p{1.95cm} p{1.95cm} p{1.6cm}}
\hline
\bfseries{ } & \bfseries{dMCS} & \bfseries{N3Sim} & \bfseries{MUCIN} & \bfseries{NanoNS} & \bfseries{BINS} & \bfseries{BNSim} \\ 
\hline
Development language & Java & Java & MATLAB & NS-2, C$++$, Tcl & Java & Java \\
Parallelization & Yes & No & No & No & No & No \\
Open source  & No  & Yes  & Yes  &  No & No & Yes\\
Propagation  & Diffusion & Diffusion & Diffusion & Diffusion & Diffusion & Bacteria \\
Track each carrier & Yes & Yes & Yes & No & Yes & Yes\\
Reception & Absorption & Sampling & Absorption & Berg, Gillespie & Receptors & Receptors \\
Imperfect reception & No & No & Yes & No & Yes & No \\
Support unbounded medium & No & Yes$^a$ & Yes & Yes & Yes & No \\
Environment dimensions & 3-D & 2-D, 3-D$^a$ & 1-D $\sim$ 3-D & 3-D & 3-D & 3-D\\
Molecule interactions & No & No & No & No & Yes & No \\
Sending consecutive symbols & No & Yes$^a$ & Yes & No & No & Yes\\
\hline
\end{tabular}
\renewcommand{\arraystretch}{1}
\end{center}
\hfill\footnotesize{$^a$ Possible only under specific conditions.}
\end{table*}
Experimental study of MC systems is very difficult because of the expensive nature of wet-lab experimentation, and the fact that performing these experiments can be very time-consuming and laborious. Therefore, previous research activities on MC have heavily relied on simulations to verify and evaluate the new communication solutions. These simulators are required to precisely track the behavior of information-carrying particles within a realistic environment. In this section, we provide an overview of some of the most recent MC simulators in the literature. Table~\ref{tb:sim_comparison} provides a comparison matrix for these simulators.

In \cite{akk14}, distributed simulations based on high level architecture (HLA) were proposed and analyzed for MC. The authors mostly focused on confined space and parallelism gain, and did not consider the effect of consecutive symbols in their simulator, dMCS. In \cite{gul10}, NanoNS simulation platform, written in C++ and Tcl, was introduced based on the well-known NS-2 simulation platform. The authors mainly focused on  channel properties and number of received molecules versus time. The multi-particle lattice automata algorithm was implemented in NanoNS, which divided the propagation medium into mesh. Thus, the exact particle position was not evaluated. In NanoNS, the reaction process that describes the molecular interaction between nodes and molecules, was modeled in three different ways, one being no reaction.

In \cite{mah14}, amplitude-based modulations were analyzed using a custom simulator, with non-absorbing receivers (i.e. information particles continued free diffusion after they reacted the receiver). In \cite{lla14}, a simulation framework called N3Sim was introduced, where the reception process at the receiver site was similar to the model in \cite{mah14} (i.e. non-absorbing receivers). 
In both of these simulators, it was assumed that the transmitters are a point source. In reality the transmitter nodes may have a volume and they could reflect the emitted molecules. N3Sim is a Java package that simulates particle diffusion in a 2-D environment, with an ongoing expansion to 3-D diffusion models (currently 3-D simulations are possible only under specific conditions \cite{n3sim_userG}). In these works, the reception process was simulated by counting the particles located within a given area close to a receiver node during each time step. N3Sim also considered the laws for handling electrostatic forces, inertial forces, and particle collisions.

A simulation tool, BINS, for nanoscale biological networks was developed for simulating information exchange at nanoscale \cite{fel12}. BINS is adaptable to any kind of information particle. In \cite{fel12}, the authors provided a comprehensive description of their simulation libraries, and demonstrated the capabilities of their simulator by modeling a section of a lymph node and the information transfer within it. Their simulator supports 3-D space, receptor dynamics including affinity, different carriers, lifetime for molecules, tracking each molecule, collisions, and mobile receiver nodes. BINS was upgraded to BINS2, which offers new features to simulate bounded environments. Using the BINS2 simulator, propagation in blood vessels was analyzed in \cite{fel13BlodVes} and \invitro{} experiments were simulated in \cite{fel13}. The authors simulated an experiment that focused on the communication between platelets and endothelium through the diffusion of nanoparticles \cite{fel13}. They verified their simulation results with experimental data.

In \cite{yilmaz2014simulationSO}, the MolecUlar CommunicatIoN (MUCIN) simulator for diffusion-based MC systems was presented. The MUCIN simulator is an end-to-end simulator that considers first hitting process for the signal reception. It supports 1-D to 3-D environments, sending consecutive symbols, imperfect molecule reception, extendable modulation, and filtering modules. The MUCIN simulator was developed in MATLAB and it is available in MATLAB file exchange central. 

In \cite{wei13}, an open source bacteria network simulator BNSim was introduced. BNSim supported parallel multi-scale stochastic modeling, chemical signaling pathways inside each bacterium, movement in 3-D environments, and chemotaxis. Similarly, in \cite{gre11} the authors proposed medium-range point-to-point MC that was based on the transport of DNA-encoded information between transmitters and receivers using bacterium. The authors presented the channel characterization and a simulator for the communication channel. The proposed communication scheme was as follows: first, a DNA message was embedded inside the bacterium cytoplasm. Then, the bacterium was released into the environment that in turn followed its natural instincts and propelled itself to a particular receiver, which was continuously releasing attractant particles. Finally, in the last step, the reception and decoding of the DNA message was preformed through exchanging genetic material.

In \cite{eck10} and \cite{far10}, a simulation environment for MC in confined environments was presented. The simulation environment, which could be used for Monte Carlo simulations, captured both diffusion-based propagation and active transport propagation using stationary kinesin and microtubule filaments. For the diffusion propagation, it was assumed that the collision of information particles with the channel walls are elastic.
\begin{figure*}[!t]
	\begin{center}
		\includegraphics[width=0.9\textwidth]{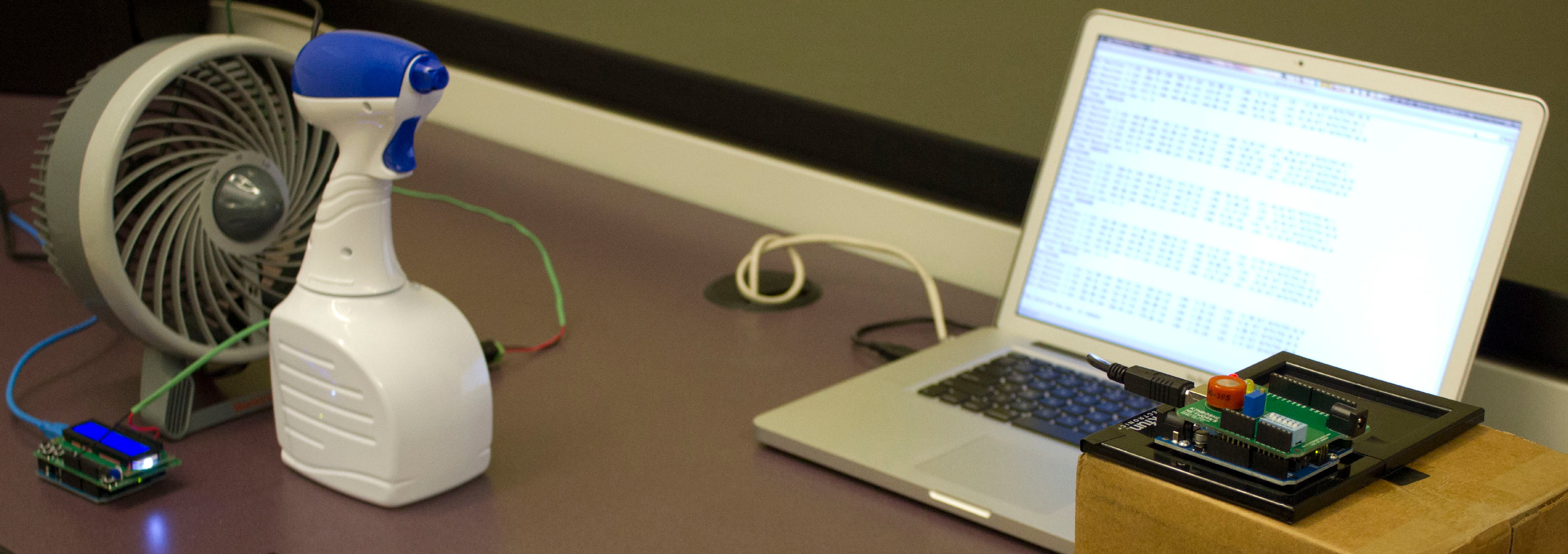}
	\end{center}
	\caption{\label{fig:tabletopMC} {The tabletop MC platform \cite{far14JSAC}.}}
\end{figure*}

\section{Practical and Experimental Systems}
\label{sec:pracSys}
As was demonstrated in the previous sections, there have been many advancements in theoretical MC over the past several years. All these advancements, however, have not translated into practical and experimental systems. Perhaps, one of the culprits is the multidisciplinary nature of MC that requires a strong collaboration between communication engineers, mechanical engineers, chemical engineers and biological engineers. Another possible hurdle is the cost associated with running wet labs.  These labs are typically very expensive to run, which makes progress slower. Despite all these issues, there have been a number of works that explore the practicality of MC in an experimental setting.

One of the first works on engineering an experimental demonstration of MC at microscales is \cite{wei01}. With the prevalence of synthetic biology and applications such as biological computation, simple intercellular communication has been used for multicellular computation \cite{tam11}. In \cite{len14}, artificial cells are used as relays (translates) to send a message to Escherichia Coli (E. Coli) cells. In this scheme, the intended message is first detected and decoded by the artificial cells, and then relayed to the E. Coli cells. A short survey on cell-to-cell communication in synthetic biology is presented in \cite{bac13}.

Most work in synthetic cell-to-cell communication consider sending a single message from a transmitter to a receiver. However, recently in \cite{ort12} it was demonstrated that a message sequence encoded in DNA can be transmitted between cells by packaging the DNA massage inside a bacteriophage. We believe that through collaborative work between communication theorist and biomedical engineers, more advanced systems could be developed in the future.

Replicating many of the experiments explained above require access to wet labs. On the other hand, the first tabletop MC system capable of transmitting short text messages over free air was developed in \cite{far13}, which does not require any access to wet labs. The macroscale platform used an alcohol metal-oxide sensor as the detector at the receiver, and an electronically controlled spray mechanism as the transmitter. Open platform Arduino microcontrollers were used to control the transmitter and the receiver. In this platform, text messages were converted into binary sequence, and the information was transmitted in a time-slotted fashion using on-off-keying, where a one was represented by spray of alcohol and a zero by no spray. The released alcohol particles propagated the environment through flow assisted propagation, which was induced using a tabletop fan. This system was also demonstrated in the IEEE International Conference on Computer Communications (INFOCOM) \cite{far14INFOCOM}. 

Figure \ref{fig:tabletopMC} shows this inexpensive platform. On the left, there is the transmitter with the spray, the fan, and Arduino for controlling the sprays. On the right is the receiver board with the sensor and the Arduino connected to a laptop. Although this system demonstrates the feasibility of MC at macroscales, it can be shrunk to microscales. For example, metal-oxide sensors can be shrunk to nanoscales \cite{rah10 , sol11} to detect various biological compounds, and a microfluidics environment could be used as the propagation channel.

An experimental platform for using MC in infrastructure monitoring was developed \cite{qiu14ICC}. In particular, the performance of MC system developed in \cite{far13} was compared with Zigbee wireless sensor nodes (operating at 2.4~GHz) inside metallic ducts. It was shown that in confined metallic environments resembling the air duct system in the buildings, the two Zigbee nodes failed to communicate as the number of bends in the ducts increase. However, it was shown that it is still possible to send limited information using the testbed developed in \cite{far13}. In \cite{wan15ICC}, a robotic platform is developed for robotic communication using persistent chemical tags.   

\begin{figure*}[t]
\centering
		\subfigure[Transmitter]
		{\includegraphics[width=0.543\textwidth,keepaspectratio]{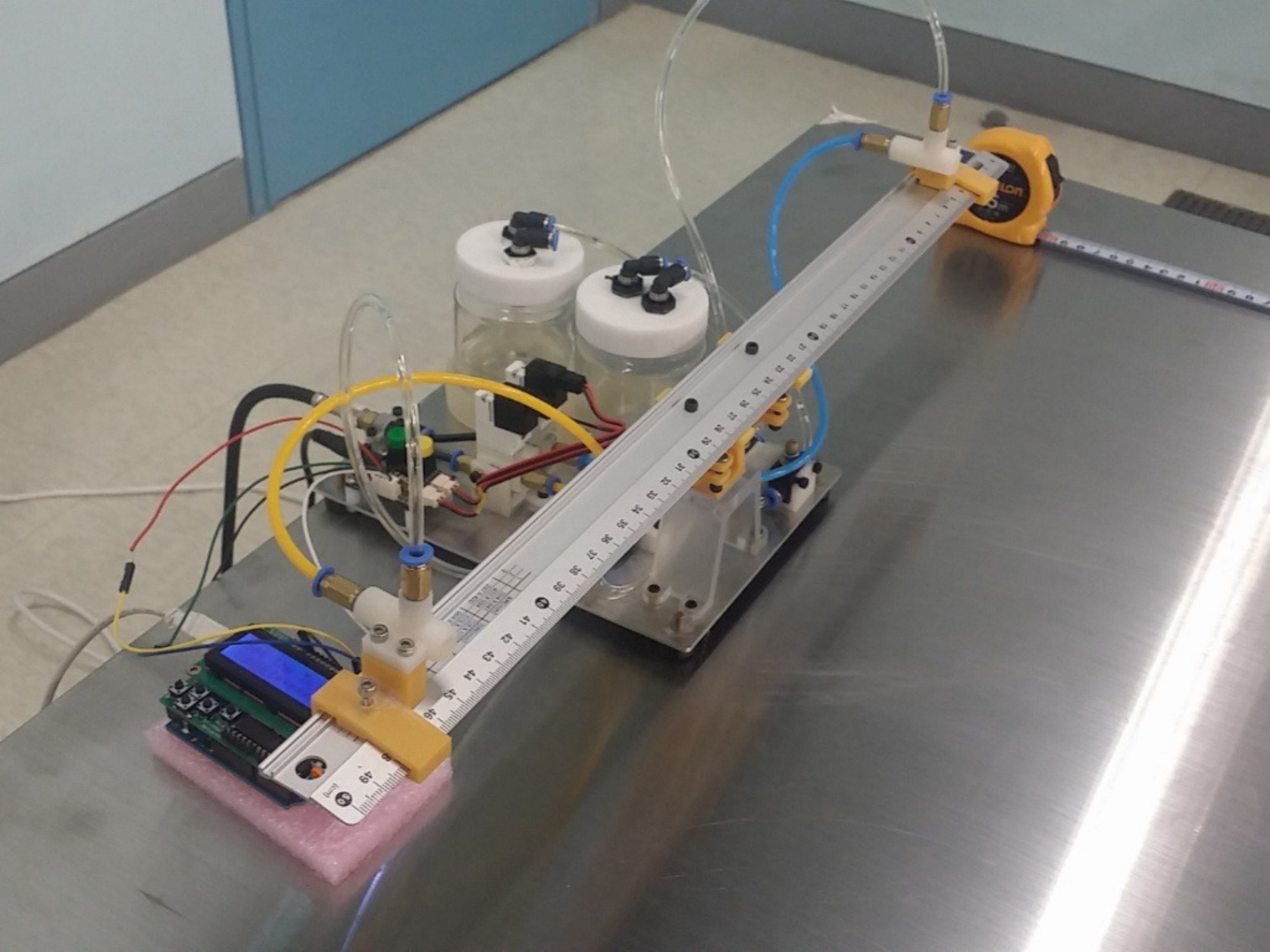}} %
		\subfigure[Receiver]
		{\includegraphics[width=0.356\textwidth,keepaspectratio]{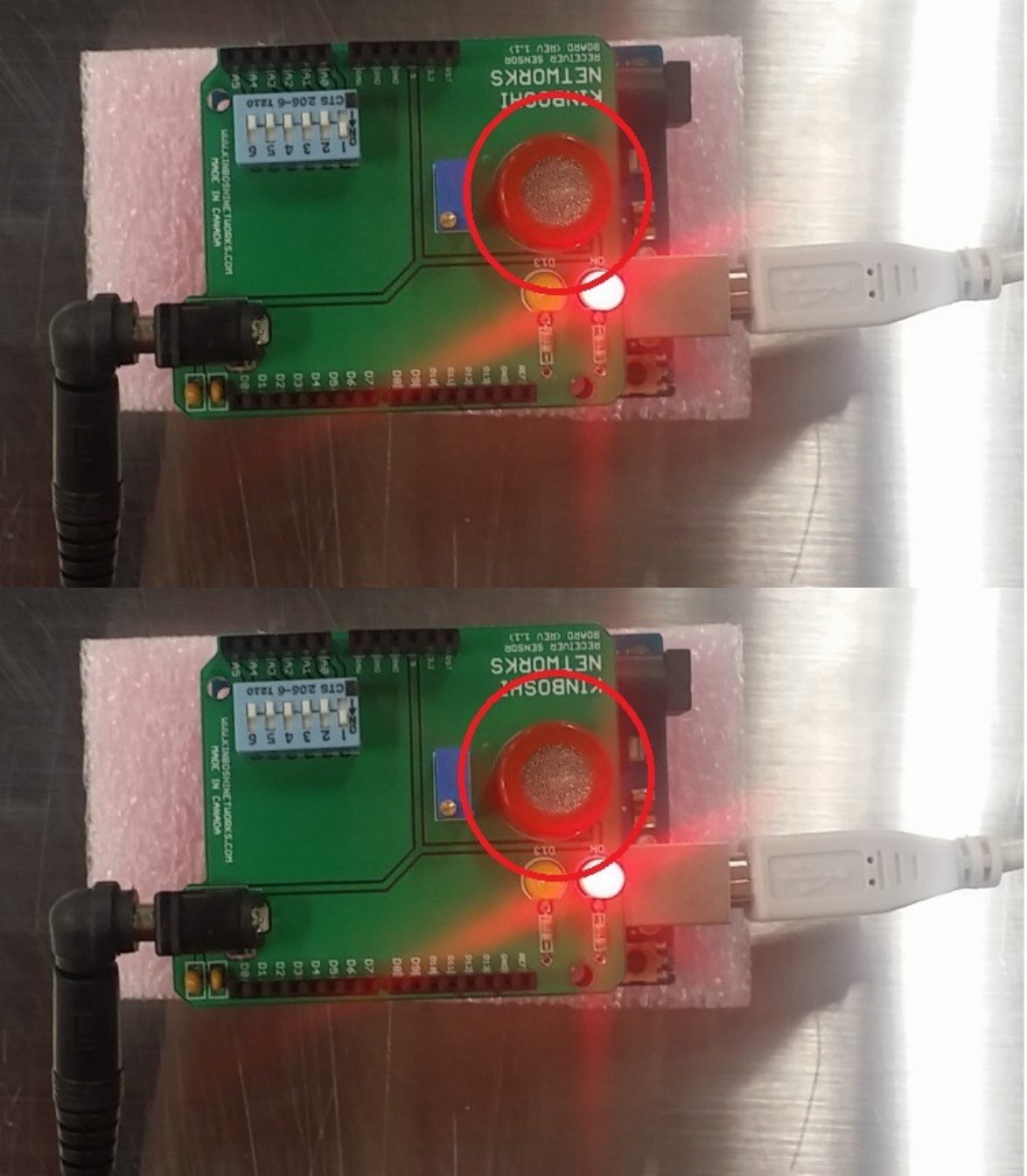}} %
	\caption{The tabletop molecular MIMO communication platform.}
	\label{Fig_test_bed_mimo}
\end{figure*}
Experimental platforms can be used to validate theoretical results and to develop more realistic models. In \cite{Kim14ICC}, it was shown that the channel response of the platform developed in \cite{far13}, did not match the theoretical results in previous works. The authors attempted to add correction factors to the previously developed theoretical models and find the best fit to the observed channel responses. In \cite{far14JSAC}, it was shown that this experimental platform tend to be nonlinear, which was in contrast to the previously derived and used models in the MC literature. The authors demonstrated that the nonlinearity may be modeled as additive Gaussian noise in certain cases. Finally, models for metal oxide sensors based MC were presented in \cite{kim15ICC}.

In~\cite{lee2015molecularMIMO_infocom,mobicom,moleMIMO}, a multiple-input-multiple-output (MIMO) MC platform was developed based on the single-input-single-output (SISO) system presented in \cite{far13}. In this device, the transmitter and the receiver were equipped with multiple sprays and sensors to further increase the data rate. The main components of this molecular MIMO system are shown in Figure~\ref{Fig_test_bed_mimo}. It was shown that the system achieves 1.78 times higher data rate compared to the SISO MC platform.  

%

\section{Technology Readiness Level \\ \& Research Challenges}
\label{sec:research_challenges}

\begin{figure*}[!t]
	\begin{center}
		\includegraphics[width=1\textwidth,keepaspectratio]{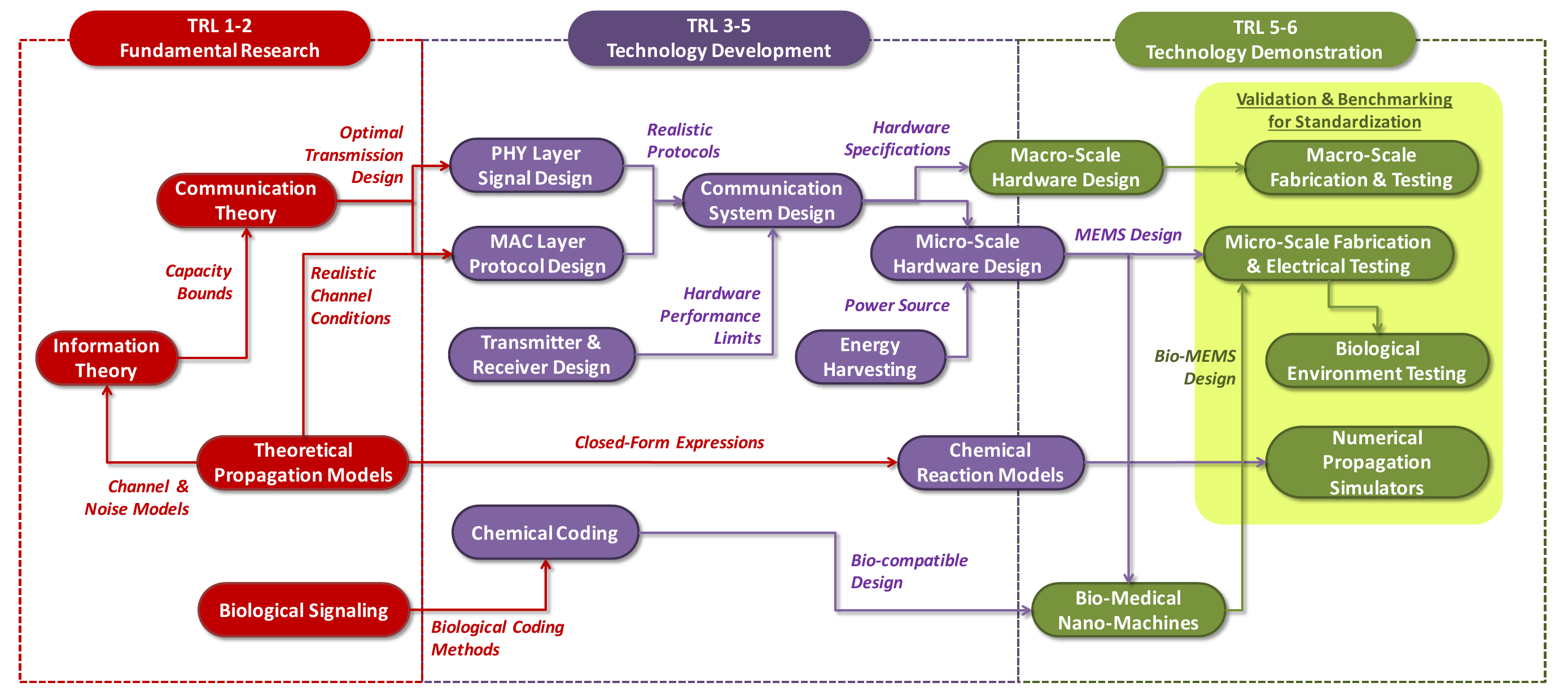}
	\end{center}
	\caption{\label{fig:TRL} A breakdown of current MC research areas in increasing order of technology readiness level. }
\end{figure*}
A missing aspect in current MC research outputs is an overall knowledge map that links singular research output areas across different technology readiness levels (TRLs). By connecting expertise across the different TRLs, a more complete scientific and engineering construct can be built, one that contributes towards standardization (i.e., IEEE 1906.1) and commercialization activities. In this section, we review the different research areas at different TRL levels, and give some examples of large-scale research projects funded by the European Commission (EU) and other funding bodies\footnote{This is by no means an exhaustive list of funded projects, but we focus on projects where there is easily accessible public information and are of a significant size and relevance to molecular communications.}.

Figure~\ref{fig:TRL} shows a breakdown of current MC research areas in increasing order of TRL. At the fundamental research level (TRL 1-2), significant work has already been conducted in the areas of: (i) information theory, (ii) communication theory, (iii) propagation modelling, and (iv) biological signal research. Together, these areas of research are able to produce beneficial research outputs in the form of capacity bounds, channel and noise models, and molecular signaling design (bio-inspired and engineered). Numerous projects around the world exist in this space, such as MINERVA (EU ERC), and Neurocommunication (EU ERC). The fundamental research contributes towards the technology development part (TRL 3-5), where most researchers currently inhabit. In here, a variety of activities exist, including the design of: PHY layer signals, MAC layer protocols, chemical coding and reactions, transmitter and receiver hardware. The design of MC protocols not only need to be aware of the optimal conditions in which capacity can be achieved, but also adverse conditions in which reliable communications maybe difficult to achieve. This is why, an understanding of realistic channel conditions is also important. The design of these vital algorithms and electronic components contribute towards an overall MC system architecture. Example projects include the PHY layer research conducted at UBC (NSERC), the SWIFT project (EU ERC), and the MoNaCo project at Georgia Tech (NSF). In order to realize a fully functional mobile system, effective power sources from energy harvesting or energy exchange needs to be built into the design. Projects such as the INFERNOS (EU FP7) produce useful outputs involving energy and information exchange at micro-scales.  The design and fabrication of macro-scale and micro-scale prototypes marks entry into the technology demonstration phase (TRL 5-6) of MCs. Existing macro-scale prototypes \cite{far13} used by researchers at York, Warwick, and Yonsei universities are crude and simplistic, but present an up-scaled research platform for experimentation. Micro-scale projects such as AtMol (EU FP7) have contributed towards MEMS fabrication of micro-/nano-scale computer chips, but few projects have made the leap towards fabricating functional MC systems (especially at the micro-scale) and this reflects the challenging nature of the research. Yet, being able to do so, and be bio-compatible and energy efficient, is currently the holy-grail of research. Only by having viable prototypes at the relevant scale, can experimentation be conducted that validate existing theoretical and numerical models developed by projects such as the CIRCLE BiNS (EU H2020) and the N3Cat N3Sim simulators. These experiments need to be conducted in a variety of environments including biological and harsh environments (such as the nano-medicine environments investigated by the NC2 team in Finland and the pipe network environments investigated by Warwick and York \cite{qiu14ICC}). The availability of hardware demonstrators and numerical modeling software will allow researchers to combine under a common scientific effort and push towards validation, benchmarking, and standardization of the technology.

Beside this general direction, we describe some of the most important open problems in MC over the next few subsections.

\subsection{Propagation Modeling}
Although variety of channel models are proposed and verified with simulations, there is a need for more realistic models and simulators. For example, many works on diffusion-based propagation assume infinite boundary conditions, which may not hold for some applications. Finally, the simulation results and the newly developed models must be validated experimentally.  The following are some open problems in propagation modeling for MC.

\begin{itemize}
	\item {\em Diffusion-based propagation}:
	\begin{itemize}
		\item {\em Models for spherical source and spherical receiver}: Although there are models for point source transmitters and spherical receiver with an infinite boundary conditions, there are no transport models between a spherical transmitter and a spherical receiver. Such models would more accurately represent applications where cell-to-cell communication is considered.
		\item {\em Confined environments}: Many previous works assume infinite boundary conditions, when considering the diffusion-based transport. However, for many applications, the communication channel may be inside a confined environment, and models based on infinite boundary conditions may not be appropriate.   
		\item {\em Heterogeneous environments}: In some medium, the diffusion coefficient can be different at various locations in the propagation channel. Previous works have not explored these environments.
		\item {\em Multiple user environments}:  The propagation models considered in the literature generally consists of single-input single-output channels, where there is only a single transmitter and a single receiver. Modeling multi-user propagation environment, where there are multiple transmitters and receivers can be very beneficial.
		\item {\em Particle decay and interaction with other particles}: There are very few works that consider transport models where the information particles could degrade over time or interact with other chemicals in the channel (e.g. enzymes). Clearly, more work is required in this promising direction.
		\item {\em Channels with flow}: Although diffusion propagation with flow has been considered previously in the literature, most works have assumed laminar flow in the direction from the transmitter to the receiver. As part of  future work time-varying flows, and turbulent flows should also be considered. 
	\end{itemize}
	\item {\em Bacteria-based propagation}: There are very few works that have considered bacterial-based propagation channels, and more work is required on these channels.
	\item {\em Motor protein-based propagation}: In the past, computer simulation have been used to model Motor protein-based propagation channels. These simulator can be improved by relaxing some of the simplifying assumptions. Moreover, the effects of ISI is typically neglected in these propagation channels, and models for ISI are required.
\end{itemize}

\subsection{Information Theoretic and System Theoretic Models}
As was discussed in the previous sections, many information theoretic and system theoretic models have been developed for MC. However, these initial models have made numerous simplifying assumptions. For example, many models assume perfect transmitters and receivers, which may not hold in practice. Also, many models assume simplistic transport models, such as infinite boundary conditions, no ISI, or constant flow velocities. As part of future work, models for the MC transmitter and receiver must be developed, and with the advancements in propagation modeling discussed in the previous subsection, more precise information theoretic and system theoretic models can be derived. Below, some of the open problems in this direction are summarized.

\begin{itemize}
	\item {\em Transmitter models}: Very few works have considered MC transmitters as part of the information and system theoretic models. This is due to lack of any models for the transmitter. In future, channels with imperfect transmitters, as well as energy, particle, and resource limited transmitters must be considered.  
	\item {\em Receiver models}: In the literature, very few works have considered the receiver as part of the system model. However, as some experimental results have demonstrated sensors could have a significant effect on the system model \cite{far14JSAC}. More work is required on modeling different sensors and detectors for MC receivers.
	\item {\em Channels with memory}: Most practical MC channels suffer from ISI, and have long memories. In order to make the formulations and models tractable, many works on information theoretic MC channels have assumed memoryless channels with no ISI. Future work must consider relaxing some of these assumptions. 
	\item {\em General capacity of MC channel}: The general capacity of the MC channel with memory, over all possible modulation schemes, is an important open problem. However, it must be noted that this is a difficult problem to be solved.
	\item {\em Capacity of MC channel with reactions}: In some MC channels, the information particles can react with each other or with other particles in the channel. Very few works have considered these effect in formulating channel capacity expressions.
	\item {\em Optimal detection and coding}: Although the capacity of some simple MC channels are known, in some cases it is not clear how this capacity can be achieved. More work is required on optimal detectors and coding for MC.
\end{itemize}

\subsection{Modulation Schemes}
Many different modulation schemes have been proposed for MC channels in the literature, and evaluated using models or simulations.  One of the biggest challenges is to validate these modulation techniques using experiments. For example, most modulations assume perfect transmitters and receivers, which may not hold in practice. It is not clear how much effect an imperfect transmitter or receiver has on the performance of the modulation scheme. Beside this main general direction, the following are some interesting open problems.

\begin{itemize}
	\item {\em Synchronization}: Some of the modulation schemes proposed for MC, would require the transmitter and the receiver to be synchronized. Some previous work have considered solving this problem, but more work is required.
	\item {\em Channel state estimation}: For some specific modulation techniques channel state information is required at the transmitter or receiver. For MC, channel state information is typically an estimate of the distance between the transmitter and the receiver or an estimate of the diffusion coefficient. 
	\item {\em ISI cancellation}: The ISI present in many MC channels has a significant effect on the performance of modulation. Therefore, ISI cancellation techniques must be employed to mitigate this issue. 
	\item {\em Higher-order modulation}: Although many different modulation techniques have been proposed for MC, there may still be possible to design novel high-order modulation schemes which can improve performance.
\end{itemize} 

\subsection{Link Layer Design}
Although many problems at the physical layer must be solved before an MC link can be established, there are some link layer problems that could be solved in conjunction. Below are some open problems in this direction. 
\begin{itemize}
	\item {\em Multiuser channels}: Many previous work, have considered modeling single user channels, and very few have studied multiuser channels. However, for many practical applications swarms of tiny devices must be able to communication with each other. 
	\item {\em Medium access control}: Just like wireless radio communication, MC environments are shared by many users. To avoid interference, medium access control  techniques are required. 
	\item {\em Addressing and routing}:  Another important problem at the link layer is addressing and routing. In a multiuser environment, there should be a mechanism that ensure messages are delivered to the correct node. 
	\item {\em Security}: Although it is not clear how much security may be important to some MC applications, there are applications that require secure communication that should not be disrupted by an adversary. Very few work have considered the problem of security over MC links.
\end{itemize}

\subsection{Experimentation}
One of the main challenges in MC is the gap between theoretical models and experimental evaluation of these models, which stems from the multidisciplinary nature of the field. Currently, theoretical models in MC are being developed by researcher from communication engineering. To validate these models, for many of the envisioned applications, wet lab experimentation is required. However, researchers from communication engineering typically do not have access to wet labs, and these experiments can only be performed by a strong collaboration with bioengineers, biologists, and chemical engineers. 

To overcome some of these limitations, as was shown in the previous chapters, two other approaches would be 1) to develop novel experimental platforms from off-the-shelf components, and 2) to develop more realistic simulators.  We believe that some of the most valuable new contributions to the field would be on validation of the models proposed in the literature---whether through collaboration with bioengineers, or developing macroscale experimental platforms, or implementing very realistic simulators. Below are some experimental platforms that if developed, would have significant effect on the advancement of the field. 

\begin{itemize}
	\item {\em MC in aqueous environments (micro- and macro-scales)}: Although experimental platforms over free-air exist, there are no MC testbeds for liquid environments. Developing such a platform at macro- or micro-scales can be very beneficial. 
	\item {\em Multiuser MC (micro- and macro-scales)}:  Experimentation on multiuser MC has not been possible until recently, because of lack of any experimental setup. However, it is now possible to run multiuser experiments at macroscale using recently developed platforms.   
	\item {\em Encoding information on the structure of molecules or particles}: A large amount of information can be encoded in the structure of molecules (e.g. DNA), which can result in a high data rate communication.  Therefore, new techniques must be developed for rapid and reliable generation of different molecular structures (e.g. different DNA sequences).
	\item {\em MC demonstrator for systems biology (microscale)}: Some of the main applications of MC are in systems biology. Therefore, an experimental platform that demonstrates MC between synthetic biological devices can be beneficial.
	\item {\em MC demonstrator for nanotechnology (microscale)}: Another area of interest is nanotechnology, which is based on inorganic matter. Demonstrating the feasibility of MC in nanotechnology using experimentation would have significant effects on both fields.
\end{itemize}

\section{Conclusions}
\label{sec:conc}

As we showed in this paper, MC is a truly multidisciplinary field of science at the intersection of communication theory, biotechnology, biology, and chemistry. Although MC is still in its infancy, there has been many advancement in developing theoretical models for MC in the communication theory society. Despite their best efforts, however, there are still many open problems that needs to be solved. Separately, there has been many advancements in the field of biotechnology, synthetic biology, and nanotechnology, where very primitive devices have been developed for experimentation. 

We believe that we are at a critical stage in the development of the field of MC, where we need to see some practical applications emerging. This will be possible with the effective collaboration of the scientists from all these different fields of science that would lead to experimental platforms that demonstrate the feasibility of MC as a technology. The goal of this survey was to present biological, chemical, and physical processes that underlay MC, and also to present a communication theoretic formulation of this problem that may not be familiar to biotechnologists. We hope that this work  motivates more collaboration between researchers from these areas of science, and fosters more experimental work in the field.

\section*{Acknowledgment}
The authors would like to thank the editors and the anonymous reviewers for their valuable comments, which have improved the quality of this survey significantly from the original submission.

\bibliographystyle{IEEEtran}
\bibliography{MolCom_YearSorted}

\begin{IEEEbiography}[{\includegraphics[width=25mm,clip,keepaspectratio]{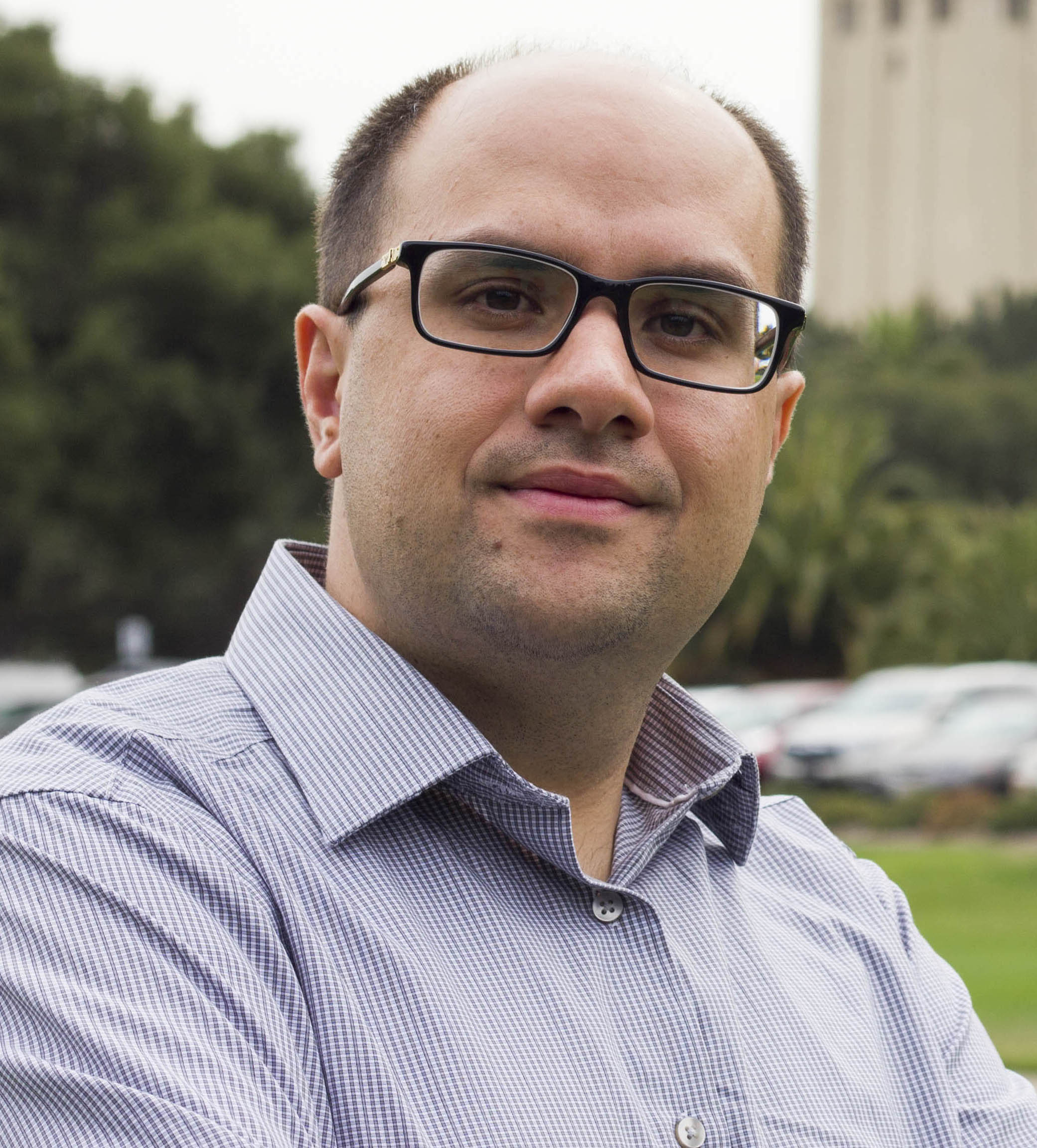}}]{Nariman Farsad}
	 received his M.Sc. and Ph.D. degrees in computer science and engineering from York University, Toronto, ON, Canada in 2010 and 2015, respectively.  He is currently a Postdoctoral Fellow with the Department of Electrical Engineering at Stanford University, where he is a recipient of Natural Sciences and Engineering Research Council of Canada (NSERC) Postdoctoral Fellowship. Nariman has won the second prize in 2014 IEEE ComSoc Student Competition: Communications Technology Changing the World, the best demo award at INFOCOM'2015, and was recognized as a finalist for the 2014 Bell Labs Prize. Nariman has been an Area Associate Editor for IEEE Journal of Selected Areas of Communication--Special Issue on Emerging Technologies in Communications, and a Technical Reviewer for a number of journals including IEEE Transactions on Signal Processing, and IEEE Transactions on Communication. He was also a member of the Technical Program Committees for the ICC'2015, BICT'2015, and GLOBCOM'2015.
\end{IEEEbiography}

\begin{IEEEbiography}[{\includegraphics[width=25mm,clip,keepaspectratio]{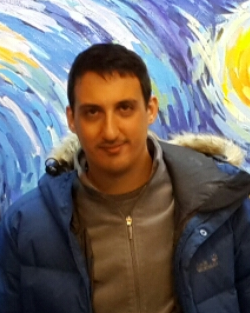}}]{H. Birkan Yilmaz}
	received his M.Sc. and Ph.D. degrees in Computer Engineering from Bogazici University in 2006 and 2012, respectively, his B.S. degree in Mathematics in 2002. Currently, he works as a post-doctoral researcher at Yonsei Institute of Convergence Technology, Yonsei University, Korea. He was awarded TUBITAK National Ph.D. Scholarship during his Ph.D. studies. He was the co-recipient of the best demo award in IEEE INFOCOM (2015) and best paper award in ISCC (2012) and AICT (2010). He is a member of TMD (Turkish Mathematical Society). His research interests include cognitive radio, spectrum sensing, molecular communications, and detection and estimation theory.
\end{IEEEbiography}

\begin{IEEEbiography}[{\includegraphics[width=25mm,clip,keepaspectratio]{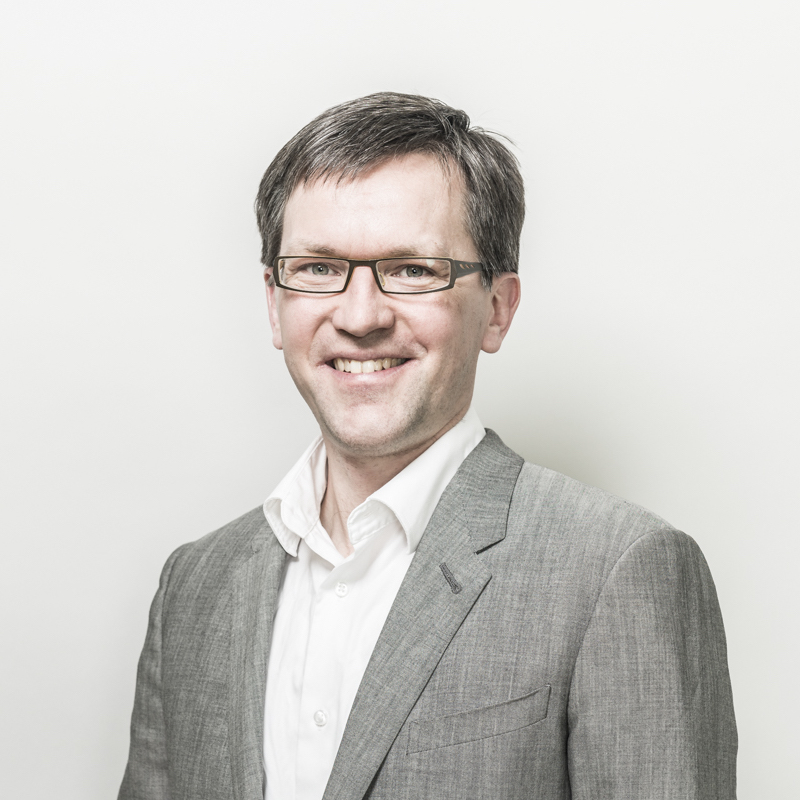}}]{Andrew Eckford}
	received the B.Eng. degree from the Royal Military College of Canada, in 1996, and the M.A.Sc. and Ph.D. degrees from the University of Toronto, in 1999 and 2004, respectively, all in electrical engineering. He is an Associate Professor in the Department of Electrical Engineering and Computer Science at York University, Toronto, Ontario. He held postdoctoral fellowships at the University of Notre Dame and the University of Toronto, prior to taking up a faculty position at York in 2006. His research interests include the application of information theory to nonconventional channels and systems, especially the use of molecular and biological means to communicate. Dr. Eckford'€™s research has been covered in media including The Economist and The Wall Street Journal. He is also a co-author of the textbook Molecular Communication, published by Cambridge University Press, and was a finalist for the 2014 Bell Labs Prize.	
\end{IEEEbiography}

\begin{IEEEbiography}[{\includegraphics[width=25mm,clip,keepaspectratio]{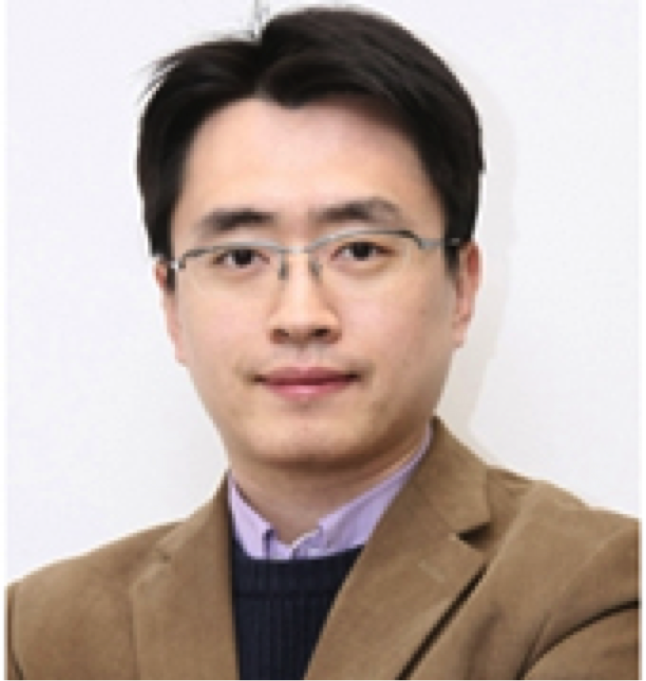}}]{Chan-Byoung Chae}
	(S'06 - M'09 - SM'12) received his Ph.D. degree  in Electrical and Computer Engineering from The University of Texas (UT), Austin, TX, USA in 2008, where he was a member of the Wireless Networking and Communications Group (WNCG). He is currently an Associate Professor in the School of Integrated Technology, College of Engineering, Yonsei University, Korea. He was a Member of Technical Staff at Bell Laboratories, Alcatel-Lucent, Murray Hill, NJ, USA from 2009 to 2011. Before joining Bell Laboratories, he was with the School of Engineering and Applied Sciences at Harvard University, Cambridge, MA, USA as a Post-Doctoral Research Fellow. 
	
	Prior to joining UT, he was a Research Engineer at the Telecommunications R\&D Center, Samsung Electronics, Suwon, Korea, from 2001 to 2005.  While having worked at Samsung, he participated in the IEEE 802.16e (mobile WiMAX) standardization, where he made several contributions and filed a number of related patents from 2004 to 2005. His current research interests include capacity analysis and interference management in energy-efficient wireless mobile networks and nano (molecular) communications. He has served/serves as an Editor for the \textsc{IEEE Trans. on Wireless Comm.}, \textsc{IEEE Trans. on Molecular, Biological, Multi-scale Comm.}, \textsc{IEEE Trans. on Smart Grid}, and \textsc{IEEE/KICS Jour. Comm. Nets}. He was a Guest Editor for the \textsc{IEEE Jour. Sel. Areas in Comm.} (special issue on molecular, biological, and multi-scale comm.). He is an IEEE Senior Member.
	
	Dr. Chae was the recipient/co-recipient of the Best Young Professor Award from the College of Engineering, Yonsei University (2015), the IEEE INFOCOM Best Demo Award (2015), the IEIE/IEEE Joint Award for Young IT Engineer of the Year (2014), the Haedong Young Scholar Award (2013), the IEEE Signal Processing Magazine Best Paper Award (2013), the IEEE ComSoc AP Outstanding Young Researcher Award (2012), the IEEE VTS Dan. E. Noble Fellowship Award (2008), the Gold Prize (1st) in the 14th/19th Humantech Paper Contests, and the KSEA-KUSCO scholarship (2007). He also received the Korea Government Fellowship (KOSEF) during his Ph.D. studies.
\end{IEEEbiography}

\begin{IEEEbiography}[{\includegraphics[width=25mm,clip,keepaspectratio]{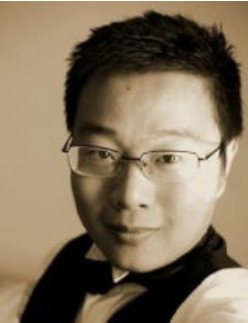}}]{Weisi Guo}  received his M.Eng., M.A. and Ph.D. degrees from the University of Cambridge. He is currently an Assistant Professor and Co-Director of Cities research theme at the School of Engineering, University of Warwick. He is the recipient of the IET Innovation Award 2015 and a finalist in the Bell Labs Prize 2014. He is a co-inventor of world's first molecular communication prototype and the author of VCEsim LTE System Simulator. He has published over 70 papers and his research interests are in the areas of heterogeneous networks, molecular communications, complex networks, green communications, and mobile data analytics.
\end{IEEEbiography}

\end{document}